\documentclass[twocolumn,english,pra]{revtex4}
\usepackage[T1]{fontenc}
\usepackage[latin9]{inputenc}
\usepackage{color}
\usepackage{float}
\usepackage{amsmath}
\usepackage{amssymb}
\usepackage{stmaryrd}
\usepackage{graphicx}

\makeatletter
\@ifundefined{textcolor}{}
{%
 \definecolor{BLACK}{gray}{0}
 \definecolor{WHITE}{gray}{1}
 \definecolor{RED}{rgb}{1,0,0}
 \definecolor{GREEN}{rgb}{0,1,0}
 \definecolor{BLUE}{rgb}{0,0,1}
 \definecolor{CYAN}{cmyk}{1,0,0,0}
 \definecolor{MAGENTA}{cmyk}{0,1,0,0}
 \definecolor{YELLOW}{cmyk}{0,0,1,0}
}

\@ifundefined{definecolor}
 {\usepackage{color}}{}
\makeatother

\makeatother

\usepackage{babel}
\begin{document}

\title{Quantifying the mesoscopic quantum coherence of approximate NOON
states and spin-squeezed two-mode Bose-Einstein condensates}

\author{B. Opanchuk, L. Rosales-Z\'arate, R. Y. Teh and M. D. Reid}

\affiliation{Centre for Quantum and Optical Science, Swinburne University of Technology,
Melbourne, Australia}
\begin{abstract}
We examine how to signify and quantify the mesoscopic quantum coherence
of approximate two-mode NOON states and spin-squeezed two-mode Bose-Einstein
condensates (BEC).  We identify two criteria that verify a nonzero
quantum coherence between states with quantum number different by
$n$. These criteria negate certain mixtures of quantum states,
thereby signifying a generalised $ $$n$-scopic Schrodinger cat-type
paradox.  The first criterion is the correlation $\langle\hat{a}^{\dagger n}\hat{b}^{n}\rangle\neq0$
(here $\hat{a}$ and $\hat{b}$ are the boson operators for each mode).
The correlation manifests as interference fringes in $n$-particle
detection probabilities and is  also measurable via quadrature phase
amplitude and spin squeezing measurements. Measurement of $\langle\hat{a}^{\dagger n}\hat{b}^{n}\rangle$
enables a quantification of the overall $n$-th order quantum coherence,
thus providing an avenue for high efficiency verification of a high-fidelity
photonic NOON states. The second criterion is based on a quantification
of the measurable spin-squeezing parameter $\xi_{N}$. We apply the
criteria to theoretical models of NOON states in lossy interferometers
and double-well trapped BECs. By analysing existing BEC experiments,
we demonstrate generalised atomic ``kitten'' states and atomic quantum
coherence with $n\gtrapprox10$ atoms.
\end{abstract}
\maketitle

\section{introduction}

In 1935 Schrodinger considered the preparation of a macroscopic system
in a quantum superposition of two macroscopically distinguishable
states \cite{Schrodinger-1}. Such systems are called ``Schrodinger
cat-states'' after Schrodinger's example of a cat in a superposition
of dead and alive states. The preparation of such states in the laboratory
is difficult due to the existence of external couplings, which cause
the superposition state to decohere to a classical mixture \cite{ystbeamsplit-1-1,decoherence}.
While for the mixture the ``cat'' is probabilistically ``dead''
or ``alive'', the paradox is that for the superposition the ``cat''
is apparently neither ``dead'' or ``alive''. Developments in quantum
optics and the cooling of atoms and mechanical oscillators have made
the generation of mesoscopic cat-states feasible \cite{cats,mirror}.
This is interesting for atomic systems where a superposition of a
massive system being in two states at different locations might be
created. \textcolor{black}{Ghirardi, Rimini, Weber} \cite{grw} and
Diosi \cite{diosi} and Penrose \cite{penrose} have proposed that
for such systems decoherence mechanisms would prevent the formation
of the Schrodinger cat superposition states. To carry out tests, firm
proposals are required for the creation and detection of cat-states.

A major consideration for cat-state experiments is that the generation
of the cat-state is not likely to be ideal. This is especially true
for larger $N$. One of the most well-studied cat-states is the NOON
state \cite{dowlingreveiw,NOON,NOONRefN5,boto,hongoumadel,macro-hong ou mandel}
\begin{equation}
|\psi_{NOON}\rangle=\frac{1}{\sqrt{2}}\{|N\rangle_{a}|0\rangle_{b}+e^{i\phi}|0\rangle_{a}|N\rangle_{b}\}\label{eq:noon-state}
\end{equation}
where $N$ particles or photons are superposed as being in the spatial
mode $a$ or the spatial mode $b$. Ideally, a cat-state requires
$N\rightarrow\infty$ but ``\emph{$N$-scopic} kitten-state'' realisations
focus on finite $N>1$. Here $|n\rangle_{a}$ ($|m\rangle_{b}$) are
the eigenstates of particle number $\hat{n}_{a}=\hat{a}^{\dagger}\hat{a}$
($\hat{n}_{b}=\hat{b}^{\dagger}\hat{b}$), respectively, and $\hat{a},$
$\hat{a}^{\dagger}$ ($\hat{b}$, $\hat{b}^{\dagger}$) are the boson
operators for mode $a$ ($b$). The NOON states have been generated
in optics for $N$ up to $5$ \cite{NOONRefN5}  and with atoms for
$N=2$ \cite{atom hongmandel}.  At low $N$ however photon detection
efficiencies are usually very low and results are often obtained by
postselection processes. 

Generating for higher $N$ is challenging. Proposals exist to exploit
the nonlinear interactions formed from Bose Einstein condensates (BEC)
trapped in the spatially separated wells of an optical lattice \cite{carr-1,cirac,twowelleprsteerNOON,wallsmil-1,murray,gordan savage,BEC NOON theory frank,carr2007,chang_bec}.
Under some conditions, theory shows that the atoms can tunnel between
the wells, resulting in the formation of a NOON superposition. However,
it is known that for realistic parameters, the states generated are
in fact of the type 
\begin{equation}
|\psi\rangle=\sum_{m=0}^{N}d_{m}|N-m\rangle_{a}|m\rangle_{b}\label{eq:sup2-1-1}
\end{equation}
where there exist nonzero probabilities for numbers other than $0$
or $N$ \cite{gordan savage,carr-1,twowelleprsteerNOON,murray,carArXiv}.
(The $d_{m}$ are probability amplitudes). Oscillation between two
BEC states with significantly different mode numbers has been experimentally
observed \cite{tunoberthaler}, presumably resulting in the formation
of a superposition of type (\ref{eq:sup2-1-1}) at intermediate times. 

A key question (raised by Leggett and Garg \cite{LG}) is how to
rigorously signify the Schrodinger cat-like property of the state
in such a non-ideal scenario. In this paper we propose quantifiable
``catness'' signatures that can be applied to nonideal NOON-type
states generated in photonic and cold atom experiments. The signatures
that we examine exclude \emph{all classical mixtures }of sufficiently
separated quantum states, so that it is possible to exclude all classical
interpretations where the ``cat'' is ``dead'' \emph{or} ``alive''
(see the Conclusion for a qualification). For the cat-system that
can be found in one of two macroscopically distinguishable states
$\rho_{D}$ and $\rho_{A}$, a rigorous signature must negate \emph{all}
mixtures of the form 
\begin{equation}
\rho_{mix}=P_{D}\rho_{D}+P_{A}\rho_{A}\label{eq:mixcatS}
\end{equation}
where $P_{D}$ and $P_{A}$ are probabilities and $P_{D}+P_{A}=1$.
In our treatment, the $\rho_{D}$ ($\rho_{A}$) are density operators
for quantum states otherwise unspecified except that they give macroscopically
distinct outcomes (``alive'' or ``dead'') for a measurement of
quantum number $\hat{n}_{a}-\hat{n}_{b}$.

The first step is to generalise this approach for the nonideal case
(\ref{eq:sup2-1-1}), where there are nonzero probabilities for obtaining
outcomes in an intermediate (``sleepy'') domain over which the cat
cannot be identified as either dead or alive. In Sections II and III,
we follow Refs. \cite{ericsqcats,cavalreidgerd,LG} and consider states
$\rho_{DS}$ and $\rho_{SA}$ that give outcomes in the combined ``dead/
sleepy'' and ``sleepy/ alive'' regions respectively. The states
have overlapping outcomes indistinguishable over a range $n$. We
explain how the negation of all mixtures of the type 
\begin{equation}
\rho_{mix}=P_{-}\rho_{DS}+P_{+}\rho_{SA}\label{eq:mixcatsda}
\end{equation}
(where $P_{-}$ and $P_{+}$ are probabilities and $P_{-}+P_{+}=1$)
will imply a generalised $n$-scopic cat-type paradox, in the sense
that the system cannot be explained by any mixture of quantum superpositions
of states different by up to $n$ quanta. It is proved that the observation
of the nonzero $n$-scopic quantum coherence term 
\begin{equation}
\langle0|\langle n|\rho|0\rangle|n\rangle\neq0\label{eq:coh}
\end{equation}
($\rho$ is the density operator) will negate all mixtures of type
(\ref{eq:mixcatsda}), thus signifying a generalised $n$-scopic
``kitten'' quantum superposition. We identify two criteria that
verify the $n$-scopic quantum coherence (\ref{eq:coh}). In a separate
paper, we examine a third criterion based on uncertainty relations
and Einstein-Podolsky-Rosen steering \cite{oursteeringnoon}.

Previous studies have proposed signatures (criteria or measures) for
mesoscopic ``cat'' states (see for instance \cite{cats,catmeasures,ystbeamsplit-1-1,ericsqcats,cavalreidgerd,frowis,nonlinearcats,svet,oursteeringnoon,murray,frowis2,vedral,frowis_3,viennahorn,carArXiv}).
These include proposals based on interference fringes, entanglement
measures, uncertainty relations, negative Wigner functions and state
fidelity. Not all of these signatures however provide a direct negation
of \emph{all} the mixtures (\ref{eq:mixcatS}), (\ref{eq:mixcatsda}).
Further, most of these studies do not address the nonideal case where
there may be a range of outcomes not binnable as either ``dead''
or ``alive''. Exceptions include the work of Refs. \cite{ericsqcats,frowis,cavalreidgerd,LG,vedral,murray,frowis2,frowis_3,macro-coh_verdral}
which (like the work of this paper) are based on the observations
of a nonzero quantum coherence. 

The first criterion that we consider  is a nonzero \emph{$n$th order
correlation} 
\begin{equation}
\langle\hat{a}^{\dagger n}\hat{b}^{n}\rangle\neq0\label{eq:moment 4}
\end{equation}
This criterion is necessary and sufficien\emph{t} for the $n$th order
quantum coherence (\ref{eq:coh}). While normally evidenced for NOON
states by fringe patterns formed from $n$-fold photon count coincidences,
we show in Section VII how this moment can also be measured for small
$N$ using highly efficient homodyne detection and Schwinger-spin
moments. In Section IV, we show that the value of the coherence $\langle\hat{a}^{\dagger n}\hat{b}^{n}\rangle$
when suitably normalised translates to an \emph{effective fidelity
measure} of the $n$-scopic ``catness'' property of the state (which
is not given directly by the state fidelity). We provide (in Section
V) a theoretical model for the NOON state with losses, thus examining
the degradation of the fidelity measure in that case. We also show
how the fidelity measure can be applied to quantify mesoscopic quantum
coherence for the case of number states $|N\rangle$ incident on a
beam splitter (the \emph{linear beam splitter} model). The quantum
coherence (5) is optimally robust with respect to losses when $n\ll N$,
but this can be achieved for high $n$.

The criterion (\ref{eq:moment 4}) was proposed by Haigh et al to
signify NOON-type superposition states created from nonlinear interactions
in two-well BECs \cite{murray}. In Section VI, we evaluate $\langle\hat{a}^{\dagger n}\hat{b}^{n}\rangle$
in dynamical regimes suitable for the formation of approximate NOON
states, using a two-mode Josephson model (the \emph{nonlinear beam}
\emph{splitter}). In Section VII, we analyse the measurement strategy
using multi-particle interferometry. For the case of $n=2,3$, the
moment (\ref{eq:moment 4}) is readily measured in terms of Schwinger
spin observables. In fact by analysing spin squeezing data reported
from the atomic BEC experiment of Esteve et al \cite{esteve}, we
infer (in Section VIII) the existence of two-atom ($n=2$) generalised
(sometimes called ``embedded'' \cite{carr-1}) kitten-states. 

The second criterion that we consider for an $n$-scopic quantum coherence
(\ref{eq:coh}) is based on \emph{spin} \emph{squeezing} \cite{SpinSqueezing,SpinSqueezingExp}.
The amount of squeezing observed for a given number of atoms $N$
is quantified by a squeeze parameter $\xi_{N}<1$ \cite{esteve,treutlein,gross,becBS}.
In Section IIIb, we apply the methods of Ref. \cite{ericsqcats} and
prove that a given measured amount of squeezing places a lower bound
of $\frac{\sqrt{N}}{\xi_{N}}$ on the value of $n$ for which the
quantum coherence $\langle0|\langle n|\rho|0\rangle|n\rangle$ is
nonzero:  
\begin{equation}
n>\frac{\sqrt{N}}{\xi_{N}}\label{eq:sqcatN}
\end{equation}
This criterion requires $\langle\hat{a}^{\dagger}\hat{b}\rangle\neq0$
and is not therefore useful to identify ideal NOON states. However
the squeezing signature (\ref{eq:sqcatN}) is very effective in confirming
a high degree of mesoscopic quantum coherence for states (\ref{eq:sup2-1-1})
where adjacent $d_{m}$ and $d_{m+1}$ are nonzero. This occurs in
systems with high losses or linear couplings. In Section IIIb we apply
this signature to published experimental data, and confirm a mesoscopic
coherence (with $n\sim10$ atoms) in two-mode BEC systems. We note
this is consistent with the recent work of Ref. \cite{frowis,vedral}
which proposes \emph{quantifiers (measures)} of mesoscopic quantum
coherence based on Fisher information and reports significant values
of atomic coherence for BEC systems.

\section{$n$-scopic quantum coherence }

We begin by considering the outcomes of observable $2\hat{J}_{Z}=(\hat{a}^{\dagger}\hat{a}-\hat{b}^{\dagger}\hat{b})$.
For the ideal NOON state (\ref{eq:noon-state}) these are $-N$
and $N$. In the limit of large $N$, we identify the two outcomes
as ``dead ($D$)'' and ``alive ($A$)'' in order to make a simplistic
analogy with the Schrodinger cat example. How does one signify the
superposition nature of the NOON state? The density matrix $\rho$
for the superposition $|\psi_{NOON}\rangle$ has nonzero off-diagonal
coherence terms $\langle0|\langle N|\rho|0\rangle|N\rangle\neq0$
that distinguish it from the classical mixture ($P_{D}$ and $P_{A}$
are probabilities, $P_{D}+P_{A}=1$) 
\begin{equation}
\rho_{mix}=P_{D}|0\rangle|N\rangle\langle N|\langle0|+P_{A}|N\rangle|0\rangle\langle0|\langle N|\label{eq:mix3-1}
\end{equation}
Thus, the detection of the nonzero coherence $\langle0|\langle N|\rho|0\rangle|N\rangle$
serves to signify an $N$-scopic cat-state in this case. 

The ultimate objective of a ``Schrodinger cat'' experiment is to
negate classical realism at a macroscopic level. The accepted definition
of \textbf{macroscopic realism} is that a system must be in a classical
mixture of two macroscopically distinguishable states \cite{LG}.
Similarly, we take as the definition of \textbf{$N$-scopic realism}
is that a system must be in a classical mixture of two states that
give predictions different by $N$ quanta. In this paper, the meaning
of\emph{ classical mixture} is in the \emph{quantum} sense only, that
the density operator $\rho$ for the system is equivalent to a classical
mixture of the two (quantum) states, as in (\ref{eq:mix3-1}).

\begin{figure}[H]
\begin{centering}
\par\end{centering}
\begin{centering}
\includegraphics{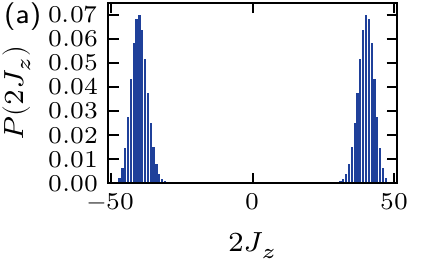}\includegraphics{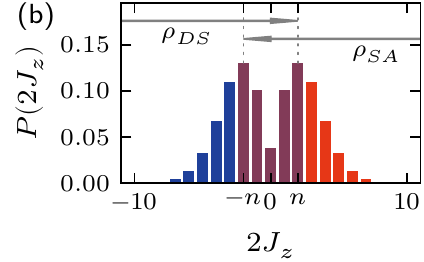} 
\par\end{centering}
\caption{\emph{Nonideal scenarios for NOON generation: }Probability $P(2j_{z})$
of an outcome of $2\hat{J_{z}}$ for the NOON state after attenuation
as modelled by a beam splitter coupling. \textcolor{black}{{} (a) $N=50$,
$\eta=0.8$ }(b) $\eta=0.05$.\textcolor{red}{{} }Similar plots are
obtained for pure NOON-type states (\ref{eq:sup2-1-1}) generated
via a Josephson two-mode interaction.  The right graph shows how
to confirm an $n$-scopic quantum coherence, as explained in the text.\label{fig:nonideal cat state-BEC-2}\textcolor{red}{}}
\end{figure}

More generally, the states generated in the experiments give outcomes
for $\hat{n}_{a}$ and $\hat{n}_{b}$ different to $0$ and $N$,
as a result of even a small amount of loss or noise in the system
(real predictions are illustrated in Figure 1). The question becomes
how to confirm by experiment that the system is indeed in a superposition
of two mesoscopically distinguishable quantum states, as opposed to
any alternative classical description where there would be no mesoscopic
``cat'' paradox. This question was examined in Ref. \cite{cavalreidgerd,ericsqcats}
for continuous outcomes realised from quadrature phase amplitude measurements
and we apply the approach given there. The following is a result
found in that paper as applied to this case. \emph{}

\textbf{\emph{Result 1$:$$-$}} \textbf{An $n$}\textbf{\emph{-scopic
quantum coherence and generalised $n$-scopic cat-paradox}}\textbf{:}
Consider the following mixture sketched in Figure 1b (for $j_{c}=0$).
 
\begin{equation}
\rho_{mix}=P_{-}\rho_{DS}+P_{+}\rho_{SA}\label{eq:mix4-1-1}
\end{equation}
where $P_{-}$ and $P_{+}$ are probabilities and $P_{-}+P_{+}=1$.
Here, $\rho_{DS}$ is a quantum state whose two-mode number state
expansion may only include eigenstates with outcome $2J_{z}<j_{c}+n$;
and $\rho_{SA}$ is a quantum state whose expansion only includes
eigenstates with $2J_{z}>j_{c}-n$. The outcome $2J_{z}\geq j_{c}+n$
is interpreted as ``alive'' and the outcome $2J_{z}\leq j_{c}-n$
is interpreted as ``dead''. The intermediate overlapping regime
is ``sleepy''. The negation of all mixtures of the type (\ref{eq:mix4-1-1})
will imply a generalised $n$-scopic ``cat-type'' paradox, in the
sense that the system cannot be viewed as either ``dead/ sleepy''
or ``alive/ sleepy''$-$ and cannot therefore be explained by any
mixture of superpositions of states different by up to $n$ quanta.
If (\ref{eq:mix4-1-1}) can be negated, then there is an \emph{$n$-scopic
generalised quantum coherence (cat-type paradox)}. The negation of
(\ref{eq:mix4-1-1}) implies that for some $n',m'$ 
\begin{equation}
_{b}\langle n+m'|_{a}\langle n'|\rho|n+n'\rangle_{a}|m'\rangle_{b}\neq0\label{eq:cohnm}
\end{equation}
(in fact $j_{c}=n'-m'$). The converse is also true. Conditions that
negate (\ref{eq:mix4-1-1}) equivalently demonstrate (\ref{eq:cohnm}),
and we refer to these conditions as signatures of $n$- scopic quantum
coherence, or of an $n$-scopic generalised cat paradox.

\emph{Proof: }The justification is that if (\ref{eq:mix4-1-1}) fails,
then the system cannot be thought of as being in one quantum state
$\rho_{1}$\emph{ or} the other $\rho_{2}$. We have not constrained
the $\rho_{1}$ or $\rho_{2}$, except to say they cannot include
both ``dead'' and ``alive'' states (each one is orthogonal to
either the dead or alive state). Hence there is a negation of the
premise that the system must always be either `` dead or alive''.
In that sense we have an analogy with the Schrodinger ``cat'' paradox
but where the ``dead'' and ``alive'' states are separated by $n$
quanta. That the coherence is nonzero follows on expanding
the density matrix in the number state basis. $\square$

We note that the nonzero $n$-scopic quantum coherence has a physical
significance, in that it is then possible (in principle) to filter
out the intermediate ``sleepy states'' using measurements of $|\hat{J}_{Z}|>n/2$
to create a conditional cat-state where the separation in $J_{Z}$
of the ``dead'' and ``alive'' states is of order $n$. This method
of preparation has been carried out experimentally \cite{macro-hong ou mandel}.
However, where the $n$-scopic coherences are small, the heralding
probability for the cat-state also becomes small, making the states
increasingly difficult to generate. This motivates Section V which
examines how to quantify the quantum coherence through experimental
signatures. First, we identify two criteria for the condition Eq.
(\ref{eq:cohnm}).

\section{Two Criteria for $n$-scopic quantum coherence}

\subsection{Correlation test}

It is well known that higher order correlations can detect NOON states.
We clarify with the following result.

\textbf{\emph{Result 2$:$$-$The $n$-th order correlation test:}}
Restricting to two-mode quantum descriptions for $\rho$, the observation
of 
\begin{equation}
\langle\hat{a}^{\dagger n}\hat{b}^{n}\rangle\neq0\label{eq:cortest}
\end{equation}
is a signature of the $n$-scopic quantum coherence (\ref{eq:cohnm}).

The Result can be proved straightforwardly by expanding the operator
$\hat{a}^{\dagger n}\hat{b}^{n}$ in terms of the Fock basis elements
$|n_{a}\rangle|n_{b}\rangle\langle m_{b}|\langle m_{a}|$ or equivalently
by considering an arbitrary density matrix $\rho$ written in the
two-mode Fock basis and noting that the condition (\ref{eq:cortest})
is equivalent to (\ref{eq:cohnm}). We will find it useful to note
the following: If the moment $\langle\hat{a}^{\dagger n}\hat{b}^{n}\rangle$
is nonzero then there is a nonzero probability that the system is
in the following\emph{ generalised $n$-scopic superposition} state:

\begin{eqnarray}
|\psi_{n}\rangle & = & a_{n'm'}^{(n)}|n'\rangle|m'+n\rangle+b_{n'm'}^{(n)}|n'+n\rangle|m'\rangle\nonumber \\
 &  & \,\,\,\,\,+d|\psi_{0}\rangle\label{eq:N sup}
\end{eqnarray}
The $a_{n'm'}^{(n)},b_{n'm'}^{(n)},d$ are probability amplitudes
satisfying $a_{n'm'}^{(n)},$$b_{n'm'}^{(n)}\neq0$, the $d$ being
unspecified. $|\psi_{0}\rangle$ is an unspecified quantum state orthogonal
to the states $|n'\rangle|m'+n\rangle$ and $|n'+n\rangle|m'\rangle$.
The meaning of ``nonzero probability that the system is in'' in
this context is that the density operator for the quantum system is
\emph{necessarily }of the form $\rho=\sum_{R}P_{R}|\psi_{R}\rangle\langle\psi_{R}|$
where at least one of the states $|\psi\rangle_{R}$ with nonzero
$P_{R}$ is an $n$-scopic superposition state $|\psi_{n}\rangle$.
 The Appendix A gives a detailed explanation of this last result.$\square$

\subsection{Spin squeezing test and application to experiment}

Significant $n$th order quantum coherence can also in some cases
be detected by observation of spin squeezing. We define the standard
Schwinger operators 
\begin{eqnarray}
\hat{J}_{X} & = & \left(\hat{a}^{\dagger}\hat{b}+\hat{a}\hat{b}^{\dagger}\right)/2\nonumber \\
\hat{j}_{Y} & = & \left(\hat{a}^{\dagger}\hat{b}-\hat{a}\hat{b}^{\dagger}\right)/(2i)\nonumber \\
\hat{J}_{Z} & = & \left(\hat{a}^{\dagger}\hat{a}-\hat{b}^{\dagger}\hat{b}\right)/2\nonumber \\
\hat{N} & = & \hat{a}^{\dagger}\hat{a}+\hat{b}^{\dagger}\hat{b}\label{eq:schspinab-1-1}
\end{eqnarray}
We consider a system described by a superposition of two-mode number
states as in (2). Thus we specify a generalised superposition as 
\begin{eqnarray}
|\psi\rangle & = & \sum_{i,j}c_{ij}|n_{i}\rangle|m_{j}\rangle\nonumber \\
 & \equiv & \sum_{k}d_{k}|\psi_{k}\rangle\label{eq:genspin}
\end{eqnarray}
 where the last line relabels (for convenience) all states of the
$ij$ array by an index $k$. In (2) we have a superposition $|\psi\rangle=\sum_{n=0}d_{n}|n\rangle|N-n\rangle$
where $N$ (the total number of particles) is fixed. This case for
large $N$ and where $d_{n}\neq0$ for some $n\neq0,N$ has been described
as a superposition of ``dead'', ``alive'' and ``sleepy'' cats.
Considering the general case (\ref{eq:genspin}), we can define for
each term $|\psi_{k}\rangle$ such that $d_{k}\neq0$ the spin number
difference $j_{k}=(n_{i}-m_{j})/2$. The aim is to put a lower bound
on the spread of possible $j_{k}$ values (depicted in Figure 1).
We define the spread as 
\begin{equation}
\delta=max\{|j_{k}-j_{k'}|\}\label{eq:maxspred}
\end{equation}
such that for $j_{k}$ and $j_{k'}$, the coefficients $d_{k}$, $d_{k'}\neq0$.
For the ideal NOON state, $\delta=N$. Here $max$ denotes the maximum
of the set. 

We can show that a certain amount of squeezing in $J_{Y}$ determines
a lower bound in the spread of eigenstates of $J_{Z}$. The method
is similar to that given in Ref. \cite{ericsqcats} which studied
quadrature phase amplitude squeezing. The spin Heisenberg uncertainty
relation is 
\begin{equation}
(\Delta\hat{J}_{Y})(\Delta\hat{J}_{Z})\geq|\langle\hat{J}_{X}\rangle|/2\label{eq:hu}
\end{equation}
Spin squeezing is obtained when\textcolor{blue}{{} }\textcolor{black}{\cite{SpinSqueezing,SpinSqueezingExp}}\textcolor{blue}{}
\begin{equation}
(\Delta\hat{J}_{Y})^{2}<|\langle\hat{J}_{X}\rangle|/2\label{eq:urspin}
\end{equation}
It is clear that in that case a low variance $(\Delta\hat{J}_{Y})^{2}$
will always imply a high variance in $\hat{J}_{Z}$. For many spin
squeezing experiments, $\langle\hat{J}_{X}\rangle\sim\langle\hat{N}\rangle/2$
which means the Bloch vector lies on the surface or near the surface
of the Bloch sphere, so that the system is close to a pure state.
Squeezing is then obtained when $(\Delta\hat{J}_{Y})^{2}<\langle\hat{N}\rangle/4$.

For pure states, the high variance in $\hat{J}_{Z}$ is associated
with a minimum spread of the superposition of eigenstates of $\hat{J}_{Z}$.
Thus, there is a lower bound on the best amount of squeezing determined
by the maximum spread (extent) $\delta$ of the superposition. In
the Appendix, following the methods of Refs. \cite{ericsqcats}, this
connection is generalised for mixed states. We prove the following
result.

\textbf{\emph{Result }}3\textbf{\emph{$:$$-$}} \textbf{\emph{Spin
squeezing test for $n$-th order quantum coherence:}} An experimentally
measured amount of spin squeezing in $J_{Y}$ is defined in terms
of a ``squeezing parameter''\textcolor{blue}{} 
\begin{equation}
\xi_{N}=\frac{\left(\Delta\hat{J}_{Y}\right)}{\sqrt{|\langle\hat{J}_{X}\rangle|/2}}\rightarrow\frac{\left(\Delta\hat{J}_{Y}\right)}{\langle N\rangle^{1/2}/2}\label{eq:sqparameter}
\end{equation}
where $\xi_{N}<1$ implies spin squeezing and $\xi_{N}=0$ is the
optimal possible squeezing (achievable as $N\rightarrow\infty$).
 Here we have taken the case where $\langle\hat{J}_{X}\rangle\sim\langle\hat{N}\rangle/2$.
We can conclude that there exists a nonzero coherence $\langle0|\langle n|\rho|0\rangle|n\rangle\neq0$
for a value $n$ where 
\begin{equation}
n>\frac{\sqrt{N}}{\xi_{N}}\label{eq:spinsqcrit}
\end{equation}
\emph{Proof: }The proof is given in the Appendix. $\square$

The particular test given by Result 3 requires $\langle\hat{J}_{X}\rangle\neq0$.
This would imply nonzero single atom coherence terms given as $\langle\hat{a}^{\dagger}\hat{b}\rangle\neq0$.
We note that the final result (\ref{eq:spinsqcrit}) indicates that
the coherence size is of order $\sqrt{N}$. Spin squeezing with a
considerable number $N$ of atoms has been observed in several atomic
experiments and excellent agreement has been obtained for $N\sim100$
with a two-mode model \cite{esteve,treutlein,gross}. Typically, the
number of atoms is $N\sim100$ or more, indicating values of quantum
coherence of order $n>10$ atoms.

\section{Measurable quantification of the mesoscopic quantum coherence}

\subsection{Catness fidelity and quantum coherence}

The observation of $\langle\hat{a}^{\dagger n}\hat{b}^{n}\rangle\neq0$
certifies the existence of the (nonzero) $n$-scopic quantum coherence,
but does not specify the \emph{magnitude} of the quantum coherence
(QC), originating from terms like
\begin{equation}
C_{n}^{(n',m')}=2|_{b}\langle n+m'|_{a}\langle n'|\rho|n+n'\rangle_{a}|m'\rangle_{b}|\label{eq:cnm}
\end{equation}
taken from Eq. (\ref{eq:cohnm}). In fact, we can easily identify
states (such as $|\alpha\rangle|\beta\rangle$) for which the $n$-scopic
quantum coherence vanishes as $n\rightarrow\infty$ (for any $m'$,
$n'$), but for which the moment $\langle\hat{a}^{\dagger n}\hat{b}^{n}\rangle$
increases. Put another way, the observation $\langle\hat{a}^{\dagger n}\hat{b}^{n}\rangle\neq0$
does not tell us the probability $P_{R}$ that the system will be
found in an associated $n$-scopic superposition Eq. (\ref{eq:N sup}),
nor the values of the probability amplitudes $a_{n'm'}^{(n)},$$b_{n'm'}^{(n)}$. 

We explain in this Section that the measured correlation $\langle\hat{a}^{\dagger n}\hat{b}^{n}\rangle$
\emph{when suitably normalised} places a lower bound on the sum of
the magnitudes of the $n$th order quantum coherences, defined as
\begin{equation}
C_{n}=\mathcal{N}\sum_{n',m'}C_{n}^{(n',m')}\label{eq:catnessfidelity}
\end{equation}
Here $\mathcal{N}$ is a normalisation factor that ensures the maximum
value of $C_{n}=1$ for the optimal case. The normalised correlation
thus gives measurable information about $C_{n}$ which is an effective
``\emph{catness-fidelity}''.

The ``catness-fidelity'' contrasts with the standard state-fidelity
measure $F$ (defined as the overlap between an experimental state
$\rho_{exp}$ and the desired superposition state \cite{fidelity}).
The standard measure is not directly sufficient to quantify a cat-state
 since it may be possible for mixtures that are not cat-type superpositions
to give a high absolute $F$ as $N\rightarrow\infty$ \cite{ystbeamsplit-1-1}.

\subsection{General Result for two-mode mixed states}

 Defining a suitable catness-fidelity is straightforward for pure
states. Any two-mode state $|\psi\rangle$ can be expanded in the
number state basis and can thus be written in terms of a superposition
of the states (\ref{eq:N sup}) but with $a_{n'm'}^{(n)}$, $b_{n'm'}^{(n)}$
arbitrary\textcolor{black}{. The state fidelity $F$ of $|\psi\rangle$
with respect to the symmetric $n$-scopic superposition 
\begin{eqnarray*}
|\psi_{sup}\rangle & = & (|n'\rangle|m'+n\rangle+e^{i\phi}|n'+n\rangle|m'\rangle)/\sqrt{2}
\end{eqnarray*}
 }\textcolor{blue}{}\textcolor{black}{is 
\begin{eqnarray}
F & = & \left|\langle\psi_{\mathrm{sup}}|\psi\rangle\right|^{2}\nonumber \\
 & = & \frac{1}{2}\left(|a_{n'm'}^{(n)}|^{2}+|b_{n'm'}^{(n)}|^{2}+2|a_{n'm'}^{(n)}b_{n'm'}^{(n)*}|\right)\label{eq:fidst}
\end{eqnarray}
where the phase $\phi$ is chosen to maxi}mise $F$. We see that
the magnitude of the quantum coherence of the pure state density operator
with respect to the states $|n'\rangle|m'+n\rangle$ and $|n'+n\rangle|m'\rangle$
is directly related to the fidelity $F$: 
\begin{eqnarray}
C_{n}^{(n',m')} & = & 2|\langle m'+n|\langle n'|\rho|n+n'\rangle|m'\rangle|\nonumber \\
 & = & 2|a_{n'm'}^{(n)}b_{n'm'}^{(n)*}|\label{eq:c}
\end{eqnarray}
\textcolor{blue}{}We note that $F=1$ if and only if $a_{n'm'}^{(n)}b_{n'm'}^{(n)*}=1/2$,
which implies $C_{n}^{(n'm')}=1$. Similarly, $C_{n}^{(n',m')}=1$
implies $F=1$. An arbitrary two-mode pure state is a superposition
of states over \emph{different} $n',m'$ and we may define as the
total ``$n$-scopic catness fidelity'' the sum of the magnitudes
of the $n$th order coherences i.e. 
\begin{eqnarray}
C_{n} & =\mathcal{N}\sum_{n',m'}C_{n}^{(n',m')}= & 2\mathcal{N}\sum_{n',m'}|a_{n'm'}^{(n)}b_{n'm'}^{(n)*}|\nonumber \\
\label{eq:purecatfidelity}
\end{eqnarray}
where $\mathcal{N}$ is a normalisation factor to ensure the maximum
value of $C_{n}=1$. A pure two-mode state with fixed $N$ as given
in the Introduction can be written
\begin{eqnarray}
|\psi\rangle & = & \sum_{m=0}^{N}d_{m}|N-m\rangle_{a}|m\rangle_{b}\nonumber \\
 & = & \sum_{m'<N/2}d_{m'+n}|m'\rangle|m'+n\rangle+d_{m}|n+m'\rangle|m'\rangle\nonumber \\
\label{eq:twomodepureN-1}
\end{eqnarray}
\textcolor{black}{(The simplification in the last line is written
for $N$ odd.) The $a_{n'm'}^{(n)}$ and $b_{n'm'}^{(n)}$ can then
be given in terms of $d_{m'+n}$ and $d_{m'}$. } For a pure state,
we see that $C_{n}$ can be inferred from the probabilities for the
mode number. However, this is not useful for the practical case of
mixtures. 

With this motivation, we note that the catness-fidelity $C_{n}$ can
be expressed in terms of the measurable higher order moments. 
\begin{eqnarray}
\langle\hat{a}^{\dagger n}\hat{b}^{n}\rangle & = & \sum_{n',m'\geq0}a_{n'm'}^{(n)}b_{n'm'}^{(n)*}\sqrt{\frac{(m'+n)!}{m'!}}\sqrt{\frac{(n'+n)!}{n'!}}\nonumber \\
\label{eq:momab-2}
\end{eqnarray}
For\textcolor{black}{{} a general two-mode state,  using that $\langle\hat{a}^{\dagger n}\hat{b}^{n}\rangle=Tr(\rho a^{\dagger n}b^{n})=\sum_{n_{a},m_{b}}\langle n_{a}|\langle m_{b}|\rho a^{\dagger n}b^{n}|n_{a}\rangle|m_{b}$,
we find }
\begin{eqnarray}
\langle\hat{a}^{\dagger n}\hat{b}^{n}\rangle & = & \sum_{n',m'\geq0}\sqrt{\frac{(n'+n)!}{n'!}}\sqrt{\frac{(m'+n)!}{(m')!}}\nonumber \\
 &  & \ \ \ \ \times\langle n'|\langle m'+n|\rho|n'+n\rangle|m'\rangle\nonumber \\
\label{eq:momab}
\end{eqnarray}
 This allows us to deduce the following general result.

\textbf{\emph{Result }}4\textbf{\emph{$:$$-$}} \textbf{\emph{Measurable
lower bound estimate to the $n$-scopic ``catness fidelity'', defined
as the sum of the magnitudes of $n$th order quantum coherences:}}

The measurable quantity
\begin{equation}
c_{n}=\frac{2|\langle\left(\hat{a}^{\dagger}\right)^{n}\hat{b}^{n}\rangle|}{S}\label{eq:catfid}
\end{equation}
gives a \emph{lower bound} to the true catness-fidelity $C_{n}$.
Here $S=sup_{n',m'}\{\sqrt{\frac{(m'+n)!}{m'!}}\sqrt{\frac{(n'+n)!}{n'!}}\}$
over values of $n'$, $m'$ satisfying that the probability $P_{m',n'+n}$
for detecting $m'$ and $n'+n$ particles in modes $b$ and $a$ is
nonzero, and also that the probability $P_{m'+n,n'}$ for detecting
$m'+n$ and $n'$ particles in modes $b$ and $a$ is nonzero. 

\emph{Proof: }The proof follows from (\ref{eq:momab}) using the definition
(\ref{eq:catnessfidelity}). $\square$ Realistically, it is difficult
in an experiment to truly verify that the probability for obtaining
a certain mode number is zero. In light of this, we deduce in the
Appendix C a correction term to the Result 4, assuming the experimentalist
is at least able to verify that the ``nonrelevant'' probabilities
$P_{m',n'+n}$, $P_{m'+n,n'}$ are sufficiently small, and that there
is a practical upper bound to the mode numbers (defined by an energy
or atom number bound).

\subsection{Ideal NOON case}

In the ideal NOON case, an experimentalist would observe $N$ particles
in mode $a$ or $N$ particles in mode $b$. Consider an experiment
where indeed only such probabilities are nonzero. This is not unrealistic
for photonic experiments with small $N$ that use postselection. The
experimentalist could deduce that the most general form of the density
operator in this case is 
\begin{equation}
\rho=P_{N}\rho_{N}+P_{alt}\rho_{alt}\label{eq:mixnoonalt-1}
\end{equation}
where $\rho_{N}$ is the density operator of a NOON superposition
(\ref{eq:N sup}) (with $n'=m'=d=0$), and $\rho_{alt}$ is an alternative
density operator describing classical mixtures of number states (namely
$|N\rangle|0\rangle$ and $|0\rangle|N\rangle$). Here, $P_{N}+P_{alt}=1$
and $P_{N}$, $P_{alt}$ are probabilities. 

We see from (\ref{eq:N sup}) that the quantity $C_{N}$ defined as
\begin{equation}
C_{N}=2|a_{00}^{(N)}b_{00}^{(N)}|P_{N}\label{eq:mixatnoon-1}
\end{equation}
 gives an \emph{effective} \emph{fidelity measure} of the state $\rho$
relative to the NOON cat state. We call the quantity $C_{N}$ the
\emph{catness-fidelity}, and note that $0\le C_{N}\le1$. Clearly,
the value of $C_{N}=1$ is optimal and can only occur if the system
$\rho$ is the pure symmetric NOON state (\ref{eq:noon-state}) for
which $|a_{00}^{(N)}|=|b_{00}^{(N)}|=\frac{1}{\sqrt{2}}$. For the
ideal NOON state, the prediction is \textcolor{green}{} $\langle a^{\dagger n}b^{n}\rangle=\delta_{Nn}N!/2$
and $S=N!$ so that the catness fidelity is indeed $1$:
\begin{equation}
c_{N}=C_{N}=\frac{2}{N!}|\langle\left(a^{\dagger}\right)^{n}b^{n}\rangle|=1\label{eq:noonid}
\end{equation}
The value of $C_{N}$ reduces for asymmetric NOON states or for mixed
states where $P_{N}<1$.

\section{Examples of quantification}

\subsection{Attenuated NOON states}

Photonic NOON states have been reported experimentally for up to $N=5$.
For a rigorous detection of a cat-like state, it is necessary to account
for losses that may arise as a result of processes including detection
inefficiencies. To model loss, we use a simple beam splitter approach
\cite{ystbeamsplit-1-1}. We calculate the moments of final detected
fields $\hat{a}_{det}$, $\hat{b}_{det}$ given by\textcolor{black}{{}
$a_{det}=\sqrt{\eta}a+\sqrt{1-\eta}a_{v},\ b_{det}=\sqrt{\eta}b+\sqrt{1-\eta}b_{v}$}
where $\hat{a}$, $\hat{b}$ are the boson operators for the incoming
field modes, prepared in a NOON state, and $a_{v}$, $b_{v}$ are
boson operators for vacuum modes associated with the environment.
Here $\eta$ is the probability that an incoming photon/ particle
is detected. We find\textcolor{red}{}\textcolor{green}{} 
\begin{equation}
\langle\hat{a}_{det}^{\dagger n}\hat{b}_{det}^{n}\rangle=\eta^{n}\langle\hat{a}^{\dagger n}\hat{b}^{n}\rangle=\eta^{n}\delta_{nN}N!/2\label{eq:atte}
\end{equation}

The system is a mixture of type $\rho=P_{N}\rho_{N}+P_{alt}\rho_{alt}$
defined in (\ref{eq:mixnoonalt-1}). The catness-fidelity signature
$C_{N}$ of Eq. (\ref{eq:mixnoonalt-1}) is measurable as $c_{N}$
($S=N!)$ defined by (\ref{eq:catfid}) and is plotted in Figure 2\textcolor{black}{.}\textcolor{red}{{}
 }Comparing with the distributions of Figure 1 which are generated
for the attenuated NOON state, we see that only the extremes $n=N$
have a nonzero coherence. As loss increases, the $N$th quantum coherence
remains (in principle) rigorously certifiable since it is predicted
that $\langle\hat{a}^{\dagger N}\hat{b}^{N}\rangle\neq0$ for all
values $\eta$. However, the fidelity $C_{N}$ is greatly reduced
with decreasing $\eta$, particularly for larger $N$ (Figure 2).

\begin{figure}[H]
\begin{centering}
\par\end{centering}
\begin{centering}
\par\end{centering}
\begin{centering}
\includegraphics{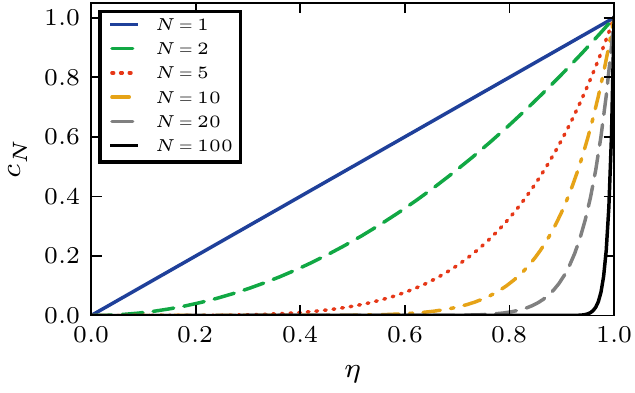}
\par\end{centering}
\caption{The $N$th order catness-fidelity $C_{N}$ (Eq. (\ref{eq:mixatnoon-1}))
for the attenuated NOON state versus detection efficiency $\eta$.
Here $C_{N}=c_{N}=2\langle\hat{a}^{\dagger n}\hat{b}^{n}\rangle/N!$\textcolor{red}{}\label{fig:nonideal cat state-BEC-1-1}\textcolor{black}{.
$c_{n}$ and $\langle\hat{a}^{\dagger n}\hat{b}^{n}\rangle=0$ for
$n<N$.}}
\end{figure}

\subsection{States formed from number states incident on a linear beam splitter }

Next we consider a two-mode number state $|N\rangle|0\rangle$ incident
at the two single-mode input ports of a beam splitter, so that $N$
quanta are incident on one arm only. The output state is the $N$-scopic
superposition (2) but with binomial coefficients: 
\begin{eqnarray}
|out\rangle & = & \sum_{m=0}^{N}d_{m}|m\rangle_{a}|N-m\rangle_{b}\,,\label{eq:bsoutputstate-1-1-2-1}
\end{eqnarray}
where $d_{m}=\sqrt{N!}/\sqrt{2^{N}m!(N-m)!}$. Different to the NOON
states, nonzero quantum coherences $\langle\hat{a}^{\dagger m}\hat{b}^{m}\rangle\neq0$
exist for all $m\leq N$\textcolor{black}{. }

\textcolor{black}{Evaluation gives that the pure state $n$-scopic
catness-fidelity (\ref{eq:purecatfidelity}) (defined as the sum of
the magnitude of all the $n$th order coherences) is 
\begin{equation}
C_{n}=\mathcal{N}_{n,N}\sum_{m=0}^{N-n}|d_{m}d_{m+n}^{*}|\label{eq:catfidn}
\end{equation}
where $\mathcal{N}_{n,N}$ is a normalisation constant to ensure the
maximum value of $C_{n}$ is $1$. For this system, the normalisation
$\mathcal{N}_{n,N}$ is determined by the bounds on the coherences
of the density matrix for a pure state. For example, where $n=N$,
$d_{0}d_{N}^{*}\leq1/2$ and hence $\mathcal{N}_{N,N}=2$. The general
results for the normalisation $\mathcal{N}_{n,N}$ are given in the
Appendix C.}\textcolor{red}{{} }\textcolor{blue}{}\textcolor{red}{}\textcolor{black}{Using
Result 4, a measurable lower bound to the catness-fidelity given by
(\ref{eq:catfid}) is
\begin{equation}
c_{n}=\frac{\mathcal{N}_{n,N}|\langle a^{\dagger n}b^{n}\rangle|}{S}\label{eq:catfidS}
\end{equation}
where $S=max\{B_{m}^{(N,n)}\}$ (for $N$ fixed) with
\begin{eqnarray*}
B_{m}^{(N,n)} & = & \sqrt{\frac{\left(m+n\right)!\left(N-m\right)!}{m!\left(N-m-n\right)!}}
\end{eqnarray*}
}The value of $m$ that gives the maximum value of $B_{m}^{(N,n)}$
is given by: $m=(N-n)/2$ \textcolor{blue}{} if $N$ and $n$ have
the same parity, and $m=(N-n\pm1)/2$ if\textcolor{red}{{} }\textcolor{black}{$n$}
and $N$ does not have the same parity.\textcolor{black}{{} }
\begin{figure}[H]
\includegraphics{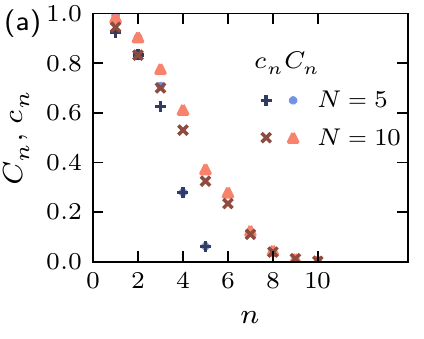}\includegraphics{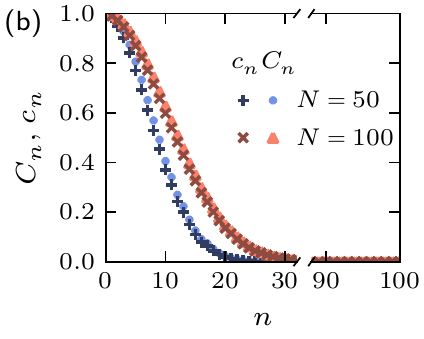}
\begin{centering}
\par\end{centering}
\begin{centering}
\par\end{centering}
\begin{centering}
\par\end{centering}
\begin{centering}
\par\end{centering}

\caption{Measures of $n$-th order quantum coherence (catness-fidelity) for
the output state of the linear beam splitter with $N$ particles incident
in one arm. $C_{N}$ and $c_{n}$ vs $n$, for $N=5$, 10 and 100.
$C_{n}\geq c_{n}$ as expected.\textcolor{blue}{}\textcolor{red}{}}
\end{figure}

\textcolor{black}{The expression $S$ can be determined from the values
of $N,m,n$ which are known for the experiment. One can then experimentally
measure the moment $\langle\hat{a}^{\dagger n}\hat{b}^{n}\rangle$
to obtain a value for $c_{n}$. The prediction is }\textcolor{blue}{
\begin{eqnarray}
{\color{black}\langle\hat{a}^{\dagger n}\hat{b}^{n}\rangle} & {\color{black}=} & {\color{black}\sum_{m=0}^{N-n}d_{m+n}^{*}d_{m}B_{m}^{(N,n)}=\frac{N!}{2^{n}(N-n)!}}\label{eq:losslessmom}
\end{eqnarray}
}\textcolor{black}{A comparison is given between the actual catness-fidelity
$C_{n}$ and the estimated one $c_{n}$ in Figure 3 for this beam
splitter case. We see that in this instance the lower bound is a good
estimate of the actual fidelity. As might be expected for this system,
the first order quantum coherence is significant whereas the highest
order coherence given by $n=N$ is small. In fact all values of fidelity
for $n>N/2$ are insignificant. We also note that for a fixed $n$,
a higher fidelity can be obtained by increasing $N$ to be much greater
than $n$.}

With attenuation present for each mode (as described in the previous
section), we evaluate the final detected moments.\textcolor{black}{{}
}The solutions are $\langle\hat{a}_{det}^{\dagger n}\hat{b}_{det}^{n}\rangle=\eta^{n}\langle\hat{a}^{\dagger n}\hat{b}^{n}\rangle$
where $\langle\hat{a}^{\dagger n}\hat{b}^{n}\rangle$ is given by
(\ref{eq:losslessmom}). The density matrix has the same dimensionality
as without losses, and the bounds on the coherences and the normalisation
$\mathcal{N}_{n,N}$ are as above. $ $Figure 4 plots the values of
the catness-fidelity $c_{n}$ versus efficiency $\eta$. We note that
the first order coherence $n=1$ is much more robust with respect
to loss, as compared to the higher order coherences. Interesting is
that for a fixed $n$, the robustness with respect to loss improves
quite dramatically if one increases the value of $N$. At high $N$,
the highest order coherences are almost immeasurable e.g. for $N=100$,
the quantum coherence becomes measurable at $n<20$. We note also
that the cut-off for a measurable $n$ increases with increasing $N$,
making generation of $n$-scopic cat-states in this generalised sense
quite feasible.

\begin{figure}[H]
\begin{centering}
\includegraphics{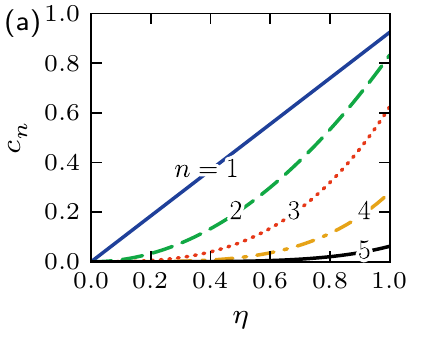}\includegraphics{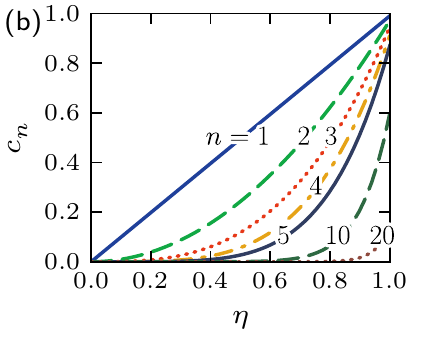}
\par\end{centering}
\caption{Measures of $n$th order quantum coherence (catness-fidelity) for
the output of the linear beam splitter versus detection efficiency
$\eta$. Definitions as for Figure 3.\textcolor{black}{{} Left $N=5$,
Right $N=100$. }\textcolor{blue}{}}
\end{figure}

\section{Mesoscopic quantum coherence in dynamical Two-well Bose-Einstein
Condensates}

\subsection{Hamiltonian and Model}

A mesoscopic NOON state can in principle be created from the nonlinear
interaction modelled by the two-mode Josephson (LMG) Hamiltonian \cite{josHam,LMGmodel}
\begin{equation}
H=\kappa\hat{a}^{\dagger}\hat{b}+\kappa\hat{b}^{\dagger}\hat{a}+\frac{g}{2}[\hat{a}^{\dagger2}\hat{a}^{2}]+\frac{g}{2}[\hat{b}^{\dagger2}\hat{b}^{2}]\label{eq:ham-1}
\end{equation}
($\hbar=1$). This Hamiltonian is well described in the literature
and models a Bose-Einstein condensate (BEC) constrained to two potential
wells of an optical lattice \cite{carr-1,cirac,twowelleprsteerNOON,wallsmil-1,murray,esteve,gordan savage,gross,twowellepr}.
The occupation of each well is modelled as a single mode (boson operators
$\hat{a}\dagger,\hat{a}$ and $\hat{b}^{\dagger},\hat{b}$ respectively).
The nonlinearity is quantified by $g$ and the tunnelling between
wells by $\kappa$. We consider a system prepared with a definite
number $N$ of atoms in one mode (well) (that denoted by $\hat{a}$).
Since the number of particles is conserved, the state at any later
time is of the form (\ref{eq:sup2-1-1}). The Hamiltonian can be
represented in matrix form and the time dependence of the $d_{m}$
solved as explained in Refs. \cite{twowellepr,carr-1,wallsmil-1,murray}.

\subsection{Two-state oscillation and creation of NOON-states}

Solutions give the probability $P(m)=|d_{m}|^{2}$ of measuring $m$
particles in the well $A$ at a given time. For some parameters, the
population oscillates between wells and there is an almost complete
transfer to the well $B$ at some tunnelling time $T_{N}$.  \textbf{}For
larger nonlinearity $g$, the system can approximate a dynamical two-state
system, showing oscillations between the two distinguishable states
$|N\rangle|0\rangle$ and $|0\rangle|N\rangle$ over long timescales
(Figure 5). At intermediate times ($\sim T_{N}/2$) before the complete
tunnelling from one state to the other, approximate NOON states can
be formed. Figure 6 depicts the probabilities $P(m)$ at the intermediate
times $T_{N}/6$ and $T_{N}/3$ that violate a Leggett-Garg inequality
\cite{LG,lg_bec}. It is known however that even for moderate $N$,
the predicted tunnelling times $T_{N}$ are typically much longer
than practical decoherence times \cite{deco,carr-1,gordan savage,atomint}.
For instance, Carr et al report impossibly long times for the typical
parameters of Rb atoms \cite{carr-1}.

\begin{figure}[H]
\begin{centering}
\includegraphics[width=0.8\columnwidth]{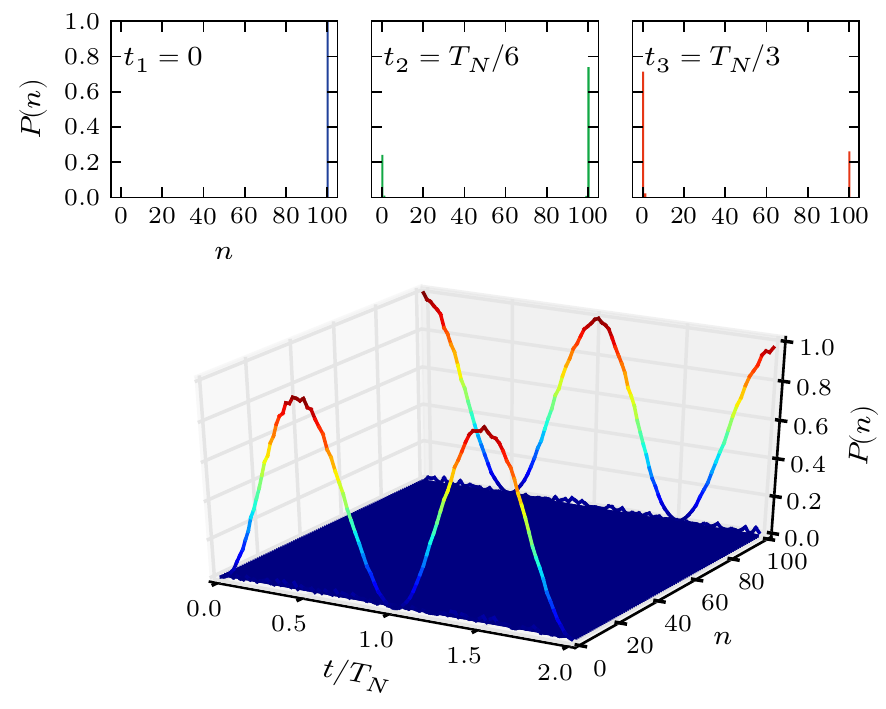}
\par\end{centering}
\begin{centering}
\par\end{centering}
\centering{}\caption{\textcolor{black}{\emph{Two-state mesoscopic dynamics: The creation
of NOON states.}}\textcolor{black}{{} Top: Probability $P(m)$ for the
number of atoms in well $a$ at times $t_{1}=0,\ t_{2}=T_{N}/6,\ t_{3}=T_{N}/3$}.
Here\textcolor{black}{{} $2j_{z}=2m-N$}. Beneath shows the two-state
oscillation.  We use $N=100$, $g=1$. Time $t$ is in units $\kappa$.
\textcolor{blue}{\label{fig:ideal-1-1-1}}\textcolor{red}{}}
\end{figure}

\begin{figure}[t]
\includegraphics{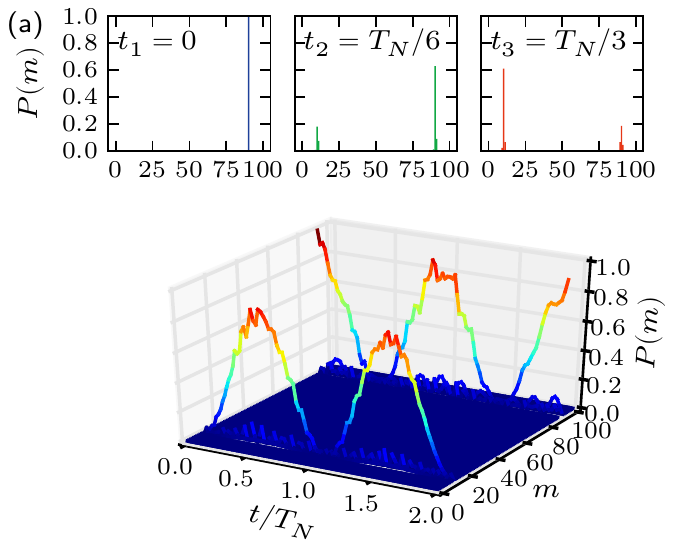}

\includegraphics{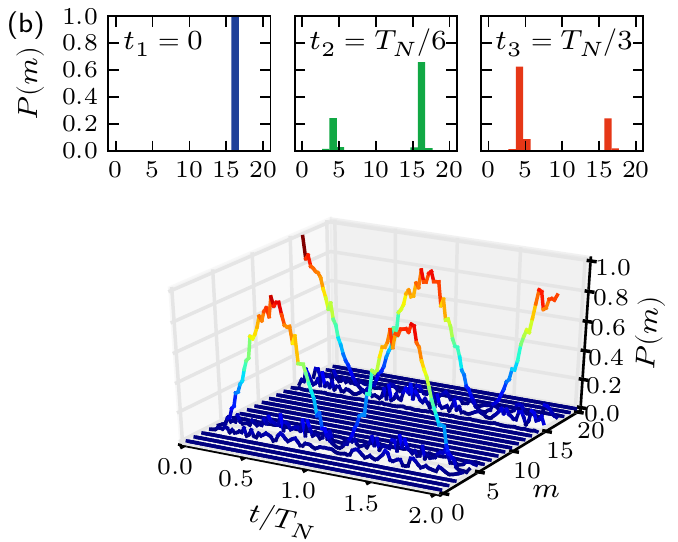}

\caption{\emph{Mesoscopic two-state oscillation and generation of NOON-type
states:} \textcolor{black}{\label{fig:nonideal N=00003D10-1-1}}\textcolor{red}{}\textcolor{black}{Top:
$N=100$, $g=2$, $n_{L}=10$.}\textcolor{black}{{} Below: $N=20$,
$g=4$, $n_{L}=4$. }\textcolor{red}{}Time $t$ is in units $\kappa$. }
\end{figure}

\begin{figure}[t]

\includegraphics{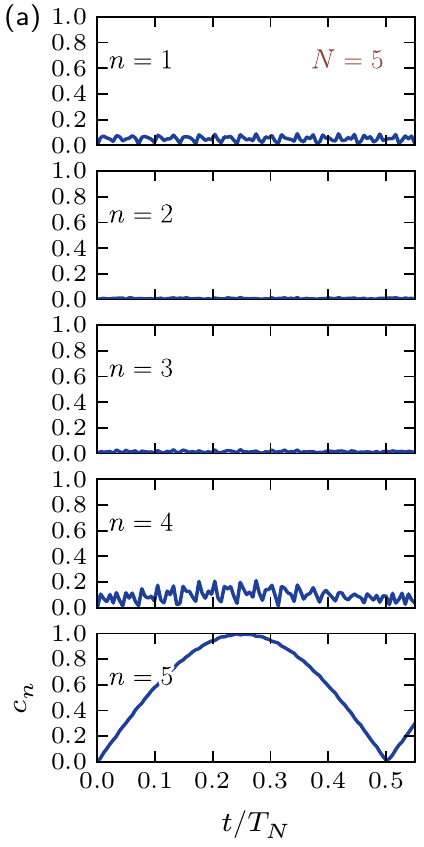}\includegraphics{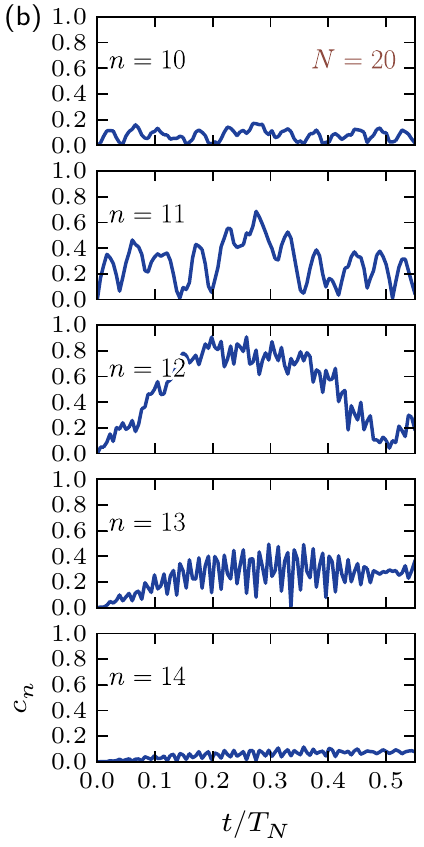}

\caption{\emph{Signifying the creation of NOON-type states under the Hamiltonian
(\ref{eq:ham-1}): }The $n$-th order quantum coherence measure $c_{n}$
versus time $t$ in units $\kappa$. Left: $N=5$, $g=10$, $n_{L}=0$.
The NOON state $N=5$ is signified by $c_{5}=1$, $c_{i}\sim0$ ($i\protect\neq5$)
at $t=T_{N}/4$. Right: $N=20$, $g=4$, $n_{L}=4$ as for Figure
6b. The large quantum coherence $c_{n}$ for $n=12$ signifies the
superposition (\ref{eq:embeddedcat}) at $t=T_{N}/4$.\textcolor{black}{\label{fig:nonideal N=00003D10}}\textcolor{red}{}\textcolor{black}{}}
\end{figure}

\begin{figure}[t]

\includegraphics{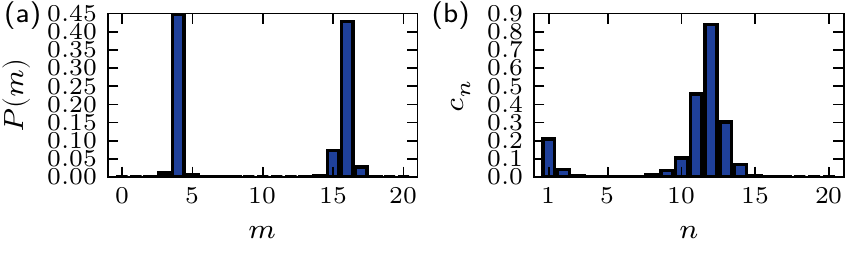}

\caption{\textcolor{black}{Plot of $P(m)$ and the $n$th order quantum coherence
$c_{n}$ for the state of Figure 6b at $t=T_{N}/4$ (as in Figure
7b).}\textcolor{red}{}}
\end{figure}

\subsection{Creation of $n$-scopic quantum superpositions}

It is possible however to generate states with a significant mesoscopic
coherence by preparing the system in an initial state $|n_{L}\rangle|N-n_{L}\rangle$
where $n_{L}\neq0,N$. As pointed out by Gordon and Savage \cite{gordan savage}
and Carr et al \cite{carr-1}, the Hamiltonian (\ref{eq:ham-1}) predicts
(in some parameter regimes) an approximate two-state oscillation between
the two states $|n_{L}\rangle|N-n_{L}\rangle$ and $|N-n_{L}\rangle|n_{L}\rangle$.
At approximately half the time for oscillation from one state to the
other, an $n$-scopic superposition state of the type given by (\ref{eq:N sup})
where $m',n'\neq0$ is formed i.e. 
\begin{equation}
|\psi\rangle=\frac{1}{\sqrt{2}}\{|n_{L}\rangle|N-n_{L}\rangle+|N-n_{L}\rangle|n_{L}\rangle\}\label{eq:embeddedcat}
\end{equation}
Here, $n=N-2n_{L}$. Such $n$-scopic superposition states have been
called ``embedded'' cat-states \cite{carr-1}. These embedded cat-states
are identical to those superpositions (\ref{eq:N sup}) discussed
in the previous section. Calculations reveal that for some parameters,
the period of oscillation reduces to practical values \cite{carr-1,gordan savage}.
Two-state oscillation of the BEC has been experimentally observed
\cite{tunoberthaler}. We present in Figure 6 predictions for this
type of oscillation with $N=20$ and $n_{L}=2$ where the solutions
indicate states (\ref{eq:embeddedcat}) with a separation of $n=16$
atoms.  \textcolor{red}{}

The question becomes how to certify the \emph{quantum coherence} of
the embedded cat-states (\ref{eq:embeddedcat}) that may be generated
in the experiment where such oscillation is observed. The value of
the catness-fidelity signature $c_{n}$ is calculated and given in
Figure 7, for the parameters of Figure 6. The $c_{n}$ for moderate
$n$ would feasibly be measurable using higher order interference
in multi-atom detection, as described in Section VII.\textbf{}

\section{Measurement of mesoscopic quantum coherence via $\langle\hat{a}^{\dagger n}\hat{b}^{n}\rangle$ }

Finally, we address how one may measure the correlation $\langle\hat{a}^{\dagger n}\hat{b}^{n}\rangle$.
The measurement of $J_{Z}$ is a photon or atom number difference,
achievable with counting detectors or imaging. Schwinger spin operators
$J_{X}=(a^{\dagger}b+b^{\dagger}a)/2,J_{Y}=\left(a^{\dagger}b-b^{\dagger}a\right)/2i$
are measured similarly as a number difference, after rotating to a
different mode pair using polarisers \cite{de m bell}; or Rabi rotations
with $\pi/2$ pulses \cite{treutlein,esteve,gross}; or beam splitters
and phase shifts. 

\subsection{Interferometric detection}

For instance, we consider the measurable output number difference
$I_{D}$ after transforming the incoming modes $ $$a$, $b$ to new
modes $c$, $d$ via a 50/ 50 beam splitter and phase shift $\varphi$:
\begin{eqnarray}
I_{D} & = & \hat{c}^{\dagger}\hat{c}-\hat{d}^{\dagger}\hat{d}\nonumber \\
 & = & \hat{a}^{\dagger}\hat{b}e^{i\phi}+\hat{a}\hat{b}^{\dagger}e^{-i\phi}\nonumber \\
 & = & 2J_{X}\cos\phi-2J_{Y}\sin\phi\label{eq:int-1}
\end{eqnarray}
Here the transformed boson operators for the new modes are $\hat{c}=(\hat{a}+\hat{b}\exp^{i\phi})/\sqrt{2}$,
$\hat{d}=(\hat{a}-\hat{b}\exp^{i\phi})/\sqrt{2}$. Selecting $\phi=0$
or $\phi=-\pi/2$ measures $J_{X}$ or $J_{Y}$. For $N=1$, $\langle\hat{a}^{\dagger}\hat{b}\rangle=\langle J_{X}+iJ_{Y}\rangle$.
The first order moment $\langle\hat{a}^{\dagger}\hat{b}\rangle$ is
thus measurable via the fringe visibility in $I_{D}$ as one varies
$\phi$ i.e. that $\langle a^{\dagger}b\rangle$ is nonzero is detectable
via first order interference. Similar transformations using atom interferometry
give the same results as explained in Ref. \cite{bryan}. If we have
a NOON state incident on the interferometer, the nonzero value of
$\langle\hat{a}^{\dagger N}\hat{b}^{N}\rangle$ can be deduced by
observation of higher order interference fringes that are signified
by an $e^{iN\phi}$ oscillation. This is the usual method for detecting
NOON states \cite{NOON,NOONRefN5,dowlingreveiw,murray}. 

The method can also be used to detect and quantify the $n$th order
quantum coherence $c_{n}$. We consider that we have a fixed total
number $N$ of particles so the input state is of the form (\ref{eq:sup2-1-1}).
The probability of detecting $N$ quanta at the output denoted by
mode $c$ is $\langle\hat{c}^{\dagger N}\hat{c}^{N}\rangle/N!$. The
probability of obtaining $M$ particles at the port $c$ is a calculable
function of the correlation functions $\langle\hat{c}^{\dagger n}\hat{c}^{n}\rangle$
where $n\geq M$. Suppose we measure $\langle\hat{c}^{\dagger n}\hat{c}^{n}\rangle$
for a given fixed $n$. Expanding we find 

\textcolor{blue}{
\begin{eqnarray}
{\color{black}\langle\hat{c}^{\dagger n}\hat{c}^{n}\rangle} & {\color{black}=} & {\color{black}\frac{1}{2^{n}}\langle(\hat{a}^{\dagger}+\hat{b}^{\dagger}e^{-i\phi})^{n}(\hat{a}+\hat{b}e^{i\phi})^{n}\rangle}\nonumber \\
{\color{black}} & {\color{black}=} & {\color{black}\frac{1}{2^{n}}\sum_{m=0}^{n}\left(\begin{array}{c}
n\\
m
\end{array}\right)\sum_{\ell=0}^{n}\left(\begin{array}{c}
n\\
\ell
\end{array}\right)}\nonumber \\
{\color{black}} & {\color{black}} & {\color{black}\times\langle(\hat{a}^{\dagger})^{n-m}(\hat{b}^{\dagger})^{m}(\hat{a})^{n-\ell}\hat{b}^{\ell}\rangle e^{i\phi(\ell-m)}}\label{eq:int-2}
\end{eqnarray}
}The terms that oscillate as $e^{in\phi}$ are proportional to the
$n$th order moment $\langle\hat{a}^{\dagger n}\hat{b}^{n}\rangle$.
Hence, if we measure ${\color{black}\langle\hat{c}^{\dagger n}\hat{c}^{n}\rangle}$,
the fringe visibility associated with this oscillation allows determination
of the magnitude of $\langle\hat{a}^{\dagger n}\hat{b}^{n}\rangle$.
Where $\langle\hat{a}^{\dagger n}\hat{b}^{n}\rangle$ is the only
nonzero moment (as for the ideal NOON states with $n=N$), only the
oscillation $e^{in\phi}$ will contribute and the higher order interference
enable a clear signature and quantification of the $n$th order quantum
coherence $\langle\hat{a}^{\dagger n}\hat{b}^{n}\rangle$. \textcolor{red}{}

For nonideal NOON states the interference method becomes less precise.
However, the rapidly oscillating terms can only arise from moments
that indicate a higher order of quantum coherence. This is evident
by the last line of (\ref{eq:int-2}). The moments are of form $\langle(\hat{a}^{\dagger})^{n-m}\hat{a}{}^{n-\ell}\hat{b}^{\dagger m}\hat{b}^{\ell}\rangle e^{i\varphi(\ell-m)}$
$ $ so the oscillation frequency where $l-m=n$ requires $l=n$ and
$m=0$ and therefore has a nonzero amplitude only if $\langle\hat{a}^{\dagger n}\hat{b}^{n}\rangle\neq0$,
which is a signature for a quantum coherence of order $n$. Similarly,
the oscillation frequency $l-m=n-1$ requires a nonzero quantum coherence
of order $n-1$. While the $\langle\hat{c}^{\dagger n}\hat{c}^{n}\rangle$
can be evaluated from the probabilities for particle counts, in practice
for large numbers $N$, resolution of atom or photon number is difficult.
Here one can measure the probability that $n$ is in a binned region
the $n$ e.g. the region $n>M$. This probability is given by \textcolor{blue}{}
\begin{eqnarray*}
P(n\geq M) & = & \sum_{n\geq M}\varsigma\langle\hat{c}^{\dagger n}\hat{c}^{n}\rangle/M!
\end{eqnarray*}
where $\varsigma$ are calculable constants. \textcolor{red}{}Here,
measurement of a nonzero amplitude for oscillations $ $$e^{iM\phi}$
with frequency $M$ or greater is evidence of quantum coherence of
order $\gtrsim M$. The high frequency oscillation can only arise
from the high order quantum coherence terms. In Figure 9, we plot
$P(n>M)$ and the Fourier analysis for the two-mode example given
in Figure 6b. 
\begin{figure}[t]

\includegraphics{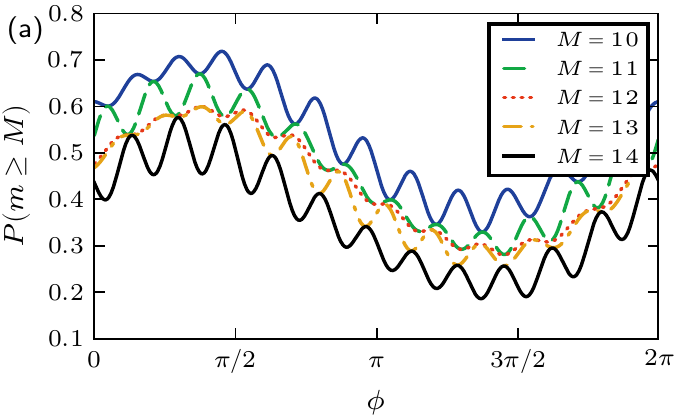}\\
\includegraphics{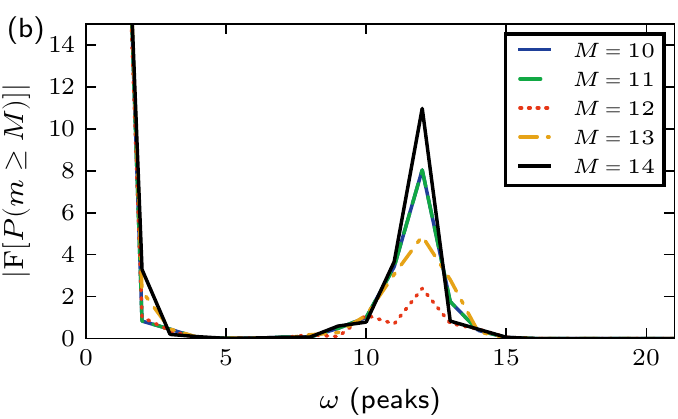}

\caption{\textcolor{red}{}(a) The probability of measuring more or equal to
$M$ photons, for $N=20$, $g=4$, $n_{L}=4$ at $t=T_{N}/4$ (same
as in Fig. 7b) after a rotation. (b) The Fourier transform of the
curves from (a), plotted against angular frequency (which is equivalent
to the number of oscillations in the range of $2\pi$), showing a
significant peak at $\omega=12$ (the expected separation of the state,
$\left(|4\rangle|16\rangle+|16\rangle|4\rangle\right)/\sqrt{2}$)
for all $M$. \textcolor{red}{}}
\end{figure}

\subsection{Spin squeezing observables and quadrature phase amplitudes}

An alternative method is given in Ref. \cite{bryan} for $N=2$. We
note that 
\begin{equation}
\langle\hat{a}^{\dagger2}\hat{b}^{2}\rangle=\langle\hat{J}_{X}^{2}\rangle-\langle\hat{J}_{Y}^{2}\rangle+i\langle\{\hat{J}_{X},\hat{J}_{Y}\}\rangle\label{eq:asecond}
\end{equation}
where $\{A,B\}\equiv AB+BA$. The real part of $\langle\hat{a}^{\dagger2}\hat{b}^{2}\rangle$
can be evaluated by measurement of $\langle\hat{J}_{X}^{2}\rangle$
and $\langle\hat{J}_{Y}^{2}\rangle$. We show that the moment is
nonzero if we can show that $\langle\hat{J}_{X}^{2}\rangle\neq\langle\hat{J}_{Y}^{2}\rangle$.
If necessary, the imaginary part can be determined by measurement
of suitably rotated spin observables defined by $\hat{J}_{\theta}=\hat{J}_{X}\cos\theta-\hat{J}_{Y}\sin\theta$.

For $N=3$ manipulation gives (see Appendix for details) \textcolor{green}{}\textcolor{blue}{}
\begin{eqnarray}
\left\langle \hat{a}^{\dagger3}\hat{b}^{3}\right\rangle  & = & 2\left\langle \hat{J}_{X}^{3}\right\rangle -\sqrt{2}(\langle\hat{J}_{\frac{\pi}{4}}^{3}\rangle+\langle\hat{J}_{\frac{3\pi}{4}}^{3}\rangle)\nonumber \\
 &  & -2i\left\langle \hat{J}_{Y}^{3}\right\rangle +i\sqrt{2}(\langle\hat{J}_{\frac{\pi}{4}}^{3}\rangle+\langle\hat{J}_{\frac{3\pi}{4}}^{3}\rangle)\label{eq:athree}
\end{eqnarray}
 where $\langle\hat{J}_{\theta}^{3}\rangle$ are measurable by standard
interferometry/ atom interferometry techniques.

We note that similar expansions can be made expressing the $a$ and
$b$ operators in terms of quadrature phase amplitudes $X$ and $P$.
For optical NOON states, this may be a useful way to accurately measure
the moments $\langle\hat{a}^{\dagger M}\hat{b}^{M}\rangle$ since
quadrature phase amplitudes can be measured with high efficiency.
Specifically, we define the amplitudes $\hat{X}$ and $\hat{P}$ by
  $\hat{a}=\hat{X}_{A}+i\hat{P}_{A}$ and $\hat{b}=\hat{X}_{B}+i\hat{P}_{B}$.
Hence (we drop the ``hats'' for convenience)
\begin{equation}
\langle\hat{a}^{\dagger}\hat{b}\rangle=\langle\hat{X}_{A}\hat{X}_{B}\rangle+\langle\hat{P}_{A}\hat{P}_{B}\rangle-i\langle\hat{P}_{A}\hat{X}_{B}+\hat{X}_{A}\hat{P}_{B}\rangle\label{eq:abquad}
\end{equation}
which is readily measurable. Continuing
\begin{eqnarray}
\langle\hat{a}^{\dagger2}\hat{b}^{2}\rangle & = & \langle(\hat{X}_{A}^{2}-\hat{P}_{A}^{2})(\hat{X}_{B}^{2}-\hat{P}_{B}^{2})\rangle+\langle\{\hat{X}_{A},\hat{P}_{A}\}\{\hat{X}_{B},\hat{P}_{B}\}\rangle\nonumber \\
 &  & -i\langle\{\hat{X}_{A},\hat{P}_{A}\}(\hat{X}_{B}^{2}-\hat{P}_{B}^{2})\rangle\nonumber \\
 &  & \,\,\,\,\,\,\,\,\,\,\,\,\,\,\,\,\,\,\,\,\,\,\,\,\,\,\,\,\,\,\,\,+i\langle(\hat{X}_{A}^{2}-\hat{P}_{A}^{2})\{\hat{X}_{B},\hat{P}_{B}\}\rangle\label{eq:absecondquad}
\end{eqnarray}
The anticommutator is measurable by rotation of the quadratures. We
define the measurable rotated quadrature phase amplitudes as $\hat{X}_{\theta}=\hat{X}\cos(\theta)+\hat{P}\sin(\theta)$
and $\hat{P}_{\theta}=-\hat{X}\sin(\theta)+\hat{P}\cos(\theta)$.
Hence, $\hat{X}_{\pi/4}=\frac{1}{\sqrt{2}}\{\hat{X}+\hat{P}\}$ and
$\hat{P}_{\pi/4}=\frac{1}{\sqrt{2}}\{-\hat{X}+\hat{P}\}$ and we note
that $\langle\hat{X}_{\pi/4}^{2}\rangle=\langle\hat{X}^{2}+\hat{P}^{2}+\hat{X}\hat{P}+\hat{P}\hat{X}\rangle/2$.
Thus, we can deduce either $\{\hat{X},\hat{P}\}$ by measuring the
moments $\langle\hat{X}^{2}\rangle$, $\langle\hat{P}^{2}\rangle$
and $\langle\hat{X}_{\pi/4}^{2}\rangle$. 

\subsection{Experimental certification of atomic quantum coherence $n\sim2$
by inferring the correlation $\langle\hat{a}^{\dagger n}\hat{b}^{n}\rangle$
from spin squeezing}

Esteve et al. experimentally realise the system modelled by the two-mode
Hamiltonian \cite{esteve}. The ground state solutions have been solved
and studied in Ref \cite{twowelleprsteerNOON}. Esteve et al report
data obtained on cooling their two-well system, including measurements
for the spin moments $\langle\hat{J}_{\theta}^{2}\rangle$ associated
with ultra-cold atomic mode populations of two wells of the optical
lattice \cite{esteve}. Their observations analyse the variances of
the Heisenberg uncertainty principle 
\begin{equation}
\Delta\hat{J}_{z}\Delta\hat{J}_{y}\geq|\langle\hat{J}_{x}\rangle|/2\sim N/4\label{eq:varuncert}
\end{equation}
They report spin squeezing in $\hat{J}_{z}$ with enhanced noise in
$\hat{J}_{y}$. They also report $\langle\hat{J}_{z}\rangle\sim0$
and $\langle\hat{J}_{y}\rangle\sim0$. Hence we can conclude 
\begin{equation}
\langle\hat{J}_{z}^{2}\rangle<N/4<\langle\hat{J}_{y}^{2}\rangle\label{eq:orderineq}
\end{equation}
Thus we deduce 
\begin{equation}
\langle\hat{J}_{y}^{2}\rangle-\langle\hat{J}_{z}^{2}\rangle\neq0\label{eq:notequal}
\end{equation}
which implies $\langle\{\hat{J}_{cx},\hat{J}_{cy}\}\rangle\neq0$
where $\hat{J}_{cx}$, $\hat{J}_{cy}$ are Schwinger operators defined
for the rotated modes $\hat{c}=(\hat{a}+\hat{b})/\sqrt{2}$ and $\hat{d}=\frac{e^{-i\pi/4}}{\sqrt{2}}(\hat{a}-\hat{b})$.
Hence we conclude 
\begin{equation}
|\langle\hat{c}^{\dagger2}\hat{d}^{2}\rangle|\neq0\label{eq:critcd}
\end{equation}
which (using the Results of Section III) gives evidence in their BEC
system of a two-atom coherence i.e. a generalised $n$-scopic superpositions
with $n=2$ of type 
\begin{equation}
|\psi_{2}\rangle=c_{20}|2\rangle_{c}|0\rangle_{d}+c_{11}|1\rangle_{c}|1\rangle_{d}+c_{02}|0\rangle_{c}|2\rangle_{d}+\psi_{0}\label{eq:supcd}
\end{equation}
where the coefficients satisfy $c_{02}\neq0$ and $c_{20}\neq0$
and where $\psi_{0}$ is orthogonal to each of $|2\rangle_{c}|0\rangle_{d}$,
$|1\rangle_{c}|1\rangle_{d}$ and $|0\rangle_{c}|2\rangle_{d}$. We
note that this is consistent with the predictions of \cite{twowelleprsteerNOON}
for the nonzero moments $\langle\hat{c}^{\dagger2}\hat{d}^{2}\rangle\neq0$
for the populations of modes $c,d$ in atomic systems with $\kappa<0$.
The observation of $\langle\hat{a}^{\dagger2}\hat{b}^{2}\rangle\neq0$
would be evidence of a superposition of atoms constrained to the modes
of the wells
\begin{equation}
|\psi_{2}\rangle=c_{20}|2\rangle_{a}|0\rangle_{b}+c_{11}|1\rangle_{a}|1\rangle_{b}+c_{02}|0\rangle_{a}|2\rangle_{b}+\psi_{0}\label{eq:supab}
\end{equation}
where $c_{02}\neq0$ and $c_{20}\neq0$. This is predicted for atomic
BEC with $\kappa>0$ \cite{twowelleprsteerNOON}. Three-atom superpositions
(for which $\langle\hat{a}^{\dagger3}\hat{b}^{3}\rangle\neq0$) and
higher are also predicted (up to $N$) and should be evident via higher
order fringe patterns, or else directly via the $J_{\theta}$ measurements
as above. 

\subsection{Entanglement}

The observation of the $n$th order quantum coherence $\langle\hat{a}^{\dagger n}\hat{b}^{n}\rangle\neq0$
is not in itself sufficient to imply entanglement\emph{. }For instance
$\psi_{0}$ in the expression (\ref{eq:supcd}) might include contributions
from terms such as $|2\rangle|2\rangle$ and $|0\rangle|0\rangle$.
This means that a separable form for $|\psi\rangle$ e.g. 
\[
|\psi\rangle=\frac{1}{2}(|2\rangle_{c}+|0\rangle_{c})(|2\rangle_{d}+|0\rangle_{d})
\]
may be possible. The separable state contrasts with the ``dead here-alive
there'' entangled superposition state whose ideal form is precisely
the NOON state e.g. for $N=2$
\[
|\psi_{2}\rangle=\frac{1}{\sqrt{2}}\{|2\rangle_{c}|0\rangle_{d}+|0\rangle_{c}|2\rangle_{d}\}
\]
In this paper, we are only concerned with how to certify an $n$-scopic
quantum superposition, without regard to entanglement. However, the
entangled case is of special interest, especially where the two modes
are spatially separated. For the ideal NOON case, we therefore point
out that one can make simple measurements to confirm the entanglement.
If one measures the individual mode numbers $n_{a}$ and $n_{b}$,
the results $0$ or $N$ are obtained for each mode. The observations
would be correlated, so that there is \emph{only }a nonzero probability
to obtain $|N\rangle|0\rangle$ or $|0\rangle|N\rangle$. This eliminates
the possibility of nonzero contributions from terms in $\psi_{0}$
and it remains only to confirm the nonzero quantum coherence in order
to confirm the entanglement. The observation of $\langle\hat{a}^{\dagger N}\hat{b}^{N}\rangle\neq0$
then becomes sufficient to certify the entanglement of the NOON state.
While simple in principle, this procedure is not so useful in practice.
For example, the attenuated NOON state of Section V would predict
a nonzero probability for obtaining $|0\rangle|0\rangle$ and a more
careful analysis is necessary to deduce entanglement. 

\section{conclusion}

We have examined how to rigorously confirm and quantify the mesoscopic
quantum coherence of non-ideal NOON states. In this paper, we link
the observation of quantum coherence to the negation of certain types
of mixtures, given as (\ref{eq:mixcatS}) and (\ref{eq:mixcatsda}).
However it is stressed we are restricting to mixtures where the ``dead''
and ``alive'' states are \emph{quantum} states that can therefore
be represented by density operators ($\rho_{A}$ and $\rho_{D}$ in
the equation (\ref{eq:mixcatS})). This contrasts with other possible
signatures of a cat-state where the dead and alive states might also
be hidden variable states, as in Ref. \cite{LG}. 

In this paper, we have focused on two criteria for the $ $$n$\emph{-th
order quantum coherence}, defined as a quantum coherence between number
states different by $n$ quanta. The first criterion is a nonzero
$n$th order moment $\langle\hat{a}^{\dagger n}\hat{b}^{n}\rangle\neq0$
and the second is a quantifiable amount of Schwinger spin squeezing.
We have shown how the first criterion can be a \emph{quantifier} of
the overall $n$th order coherence. The second criterion can be a
robust and effective signature for large $n$, and can verify high
orders of coherence in existing atomic experiments, but does not signify
all cases of $n$-scopic quantum coherence. In Sections V-VI, we have
illustrated the use of the criteria with the examples of attenuated
NOON states, number states $|N\rangle$ that pass through beam spitters,
and approximate NOON states formed from $N$ particles via nonlinear
interactions. These examples model recent photonic and atomic BEC
experiments. 

In Section VII, we have examined how the moments $\langle\hat{a}^{\dagger n}\hat{b}^{n}\rangle$
might be measured. Optical NOON states are normally verified by $n$th
order interference fringes, which imply $\langle\hat{a}^{\dagger n}\hat{b}^{n}\rangle\neq0$.
Direct photon detection normally introduces high losses which creates
low fidelities that may make significant statistics difficult, except
with post selection. We suggest that to obtain higher cat-fidelities
the moments can be measured via high efficiency quadrature phase amplitude
detection. 

Finally, in Section VIII, we analyse data from experiments, noting
that the signatures do not directly prove entanglement i.e. do not
distinguish between a local superpositions of type $|N\rangle+|0\rangle$
for one mode, and the entangled superposition of the NOON state. Hence
we cannot conclude a superposition of states with different mass locations,
although we believe this could be possible using for instance the
entanglement criteria presented in Refs. \cite{bryan,murray,hillzub}.

\section*{Acknowledgements}

We thank P. Drummond, B. Dalton, Q. He and those at the 2016 Heraeus
Seminar on Macroscopic Entanglement for discussions on topics related
to this paper. We are grateful to the Australian Research Council
for support through its Discovery Projects program. 

\appendix

\section{Result 2\label{sec:AppendixProofs}}

To explain the connection between the condition $\langle\hat{a}^{\dagger n}\hat{b}^{n}\rangle\neq0$
and the superposition state (\ref{eq:N sup}) in detail, consider
the most general two-mode quantum state for this two-mode system that
cannot be a superposition of two states distinct by $n$ quanta. We
note that any pure state $|\psi_{ent}\rangle$ can be expanded in
the two-mode number (Fock) state basis: 
\begin{eqnarray*}
|\psi_{ent}\rangle & = & \sum_{n,m}c_{nm}|n_{a}\rangle|m_{b}\rangle\\
 & = & c_{00}|0\rangle|0\rangle+c_{01}|0\rangle|1\rangle+c_{10}|1\rangle|0\rangle\\
 &  & +c_{11}|1\rangle|1\rangle+c_{12}|1\rangle|2\rangle+c_{21}|2\rangle|1\rangle\\
 &  & +c_{02}|0\rangle|2\rangle+c_{20}|2\rangle|0\rangle+...
\end{eqnarray*}
We see that if $\langle\hat{a}^{\dagger n}\hat{b}^{n}\rangle\neq0$,
then the state is necessarily of the form (\ref{eq:N sup})which
involves a superposition of two states distinct by $n$ quanta. We
note that when $\langle\hat{a}^{\dagger n}\hat{b}^{n}\rangle\neq0$,
the density operator $\rho$ for the system cannot be written in an
alternative form except to provide a nonzero coherence (\ref{eq:cohnm})
between states $|n'\rangle|m'+n\rangle$ and $|n'+n\rangle|m'\rangle$.
We conclude that the diagonal elements $_{b}\langle m'+n|_{a}\langle n'|\rho|n'\rangle_{a}|m'+n\rangle_{b}$
and $_{b}\langle m'|_{a}\langle n'+n|\rho|n'+n\rangle_{a}|m'\rangle_{b}$
are also nonzero. Thus, there is a nonzero probability $P_{D}$ that
the system is found in state $|n'\rangle|m'+n\rangle$ (that we call
``dead'') and also a nonzero probability $P_{A}$ that the system
is found in state $|n'+n\rangle|m'\rangle$ (that we call ``alive'').
Yet, the superposition state (\ref{eq:N sup}) cannot be given as
a classical mixture (\ref{eq:mix4-1-1}) which has a zero coherence
between the states $|n'\rangle|m'+n\rangle$ and $|n'+n\rangle|m'\rangle$
whose $2J_{z}$ values are different by $n$.

\section{Proof of Result 3 for Spin squeezing test}

We follow from the main text and generalise to consider a two-mode
description of the state as given by for a mixed state by a density
operator $\rho$. We expand in terms of pure states $|\psi_{R}\rangle$
so that $\rho=\sum_{R}P_{R}|\psi_{R}\rangle\langle\psi_{R}|$ for
some probabilities $P_{R}$. Each pure state $|\psi_{R}\rangle$ can
be expressed as a superposition of number eigenstates given by (\ref{eq:genspin}).
We know that the variance $ $$(\Delta\hat{J}_{Y})^{2}$ of any mixture
satisfies $(\Delta\hat{J}_{Y})^{2}\geq\sum_{R}P_{R}(\Delta\hat{J}_{Y})_{R}^{2}$.
Thus 
\begin{eqnarray*}
(\Delta\hat{J}_{Y})^{2} & \geq & \sum_{R}P_{R}(\Delta\hat{J}_{Y})_{R}^{2}\geq\sum_{R}P_{R}\frac{|\langle\hat{J}_{X}\rangle_{R}|^{2}}{4(\Delta\hat{J}_{Z})_{R}^{2}}
\end{eqnarray*}
For all the possible mixtures denoted by a choice of set $\{|\psi_{R}\rangle\}$
(where $P_{R}\neq0$) we can determine the spread $\delta_{R}$ for
each state $|\psi_{R}\rangle$ and then select the maximum of the
set $\delta_{R}$ and call it $\delta_{0}$. We select the mixture
set consistent with the density operator that has the minimum possible
value of $\delta_{0}$: That is, we determine that the density operator
cannot be expanded in a set $|\psi_{R}\rangle$ with a smaller $\delta_{0}$.
Then for the pure states of this set $|\psi_{R}\rangle$, the maximum
variance in $\hat{J}_{Z}$ is $(\Delta\hat{J}_{Z})^{2}=\delta_{0}^{2}/4$
i.e. $(\Delta\hat{J}_{Z})_{R}^{2}\leq\delta_{0}^{2}/4$. Then we see
that the uncertainty relation (\ref{eq:hu}) implies a minimum value
for the variance in $\hat{J}_{Y}$: 
\begin{eqnarray*}
(\Delta\hat{J}_{Y})_{R}^{2} & \geq\frac{|\langle\hat{J}_{X}\rangle_{R}|^{2}}{4(\Delta J_{Z})_{R}^{2}} & \geq\frac{1}{\delta_{0}^{2}}|\langle\hat{J}_{X}\rangle_{R}|^{2}
\end{eqnarray*}
Simplification gives 
\begin{eqnarray*}
(\Delta\hat{J}_{Y})^{2} & \geq & \frac{1}{\delta_{0}^{2}}\sum_{R}P_{R}|\langle\hat{J}_{X}\rangle_{R}|^{2}\geq\frac{1}{\delta_{0}^{2}}|\sum_{R}P_{R}\langle\hat{J}_{X}\rangle_{R}|^{2}\\
 & = & \frac{1}{\delta_{0}^{2}}|\langle\hat{J}_{X}\rangle|^{2}
\end{eqnarray*}
Taking the case of the spin squeezing experiments where measurements
give $\langle\hat{J}_{X}\rangle\sim\langle N\rangle/2$, we see that
\begin{equation}
(\Delta\hat{J}_{Y})^{2}\geq\frac{1}{\delta_{0}^{2}}|\langle\hat{J}_{X}\rangle|^{2}=\frac{\langle N\rangle^{2}}{4\delta_{0}^{2}}\label{eq:cond}
\end{equation}
Thus there is a lower bound on the best amount of squeezing determined
by the maximum spread (extent) $\delta_{0}$ of the superposition.
We can now prove the Result 3: The measured amount of squeezing places
a lower bound on the extent $\delta_{0}$ of the broadest superposition:
Thus if the measured squeezing is $\xi_{N}$, then from (\ref{eq:cond})
the underlying state has a minimum breadth $\delta_{0}$ of superposition
(in the eigenstates of $\hat{J}_{Z}$) given by $\delta_{0}>\frac{\langle N\rangle}{2(\Delta J_{Y})}=\frac{\sqrt{N}}{\xi_{N}}$.
The width $\delta_{0}$ of the superposition gives the extent or size
of the coherence i.e. the value of $n$ in the expression (\ref{eq:N sup}).

\section{Catness-fidelity quantifier for mixed states \label{sec:Appendix_CalculationMomentsN3-2}}

\emph{Discussion in terms of superposition states: }We give the proof
of Result (4) in terms of the superposition states. The experiment
may confirm a range of values of $j_{z}$ for $J_{z}$ for which $\langle\hat{a}^{\dagger2j_{z}}\hat{b}^{2j_{z}}\rangle\neq0$.
Take one such value: $2j_{z}=N_{0}$. Then we know there is a nonzero
probability $P_{N_{0}}$ that the system be in a superposition of
form 
\begin{eqnarray}
|\psi_{N_{0}}\rangle_{nm} & = & a_{N_{0}}^{(n,m)}|n\rangle|m+N_{0}\rangle+b_{N_{0}}^{(n,m)}|n+N_{0}\rangle|m\rangle\nonumber \\
 &  & \,\,+c|\psi\rangle\label{eq:state6-1-2}
\end{eqnarray}
where $a_{N_{0}}^{(n,m)},$$b_{N_{0}}^{(n,m)}\neq0$. Based on the
measured moments, we can write the density operator in the general
form 
\begin{equation}
\rho=\sum_{n,m}P_{N_{0}}^{(n,m)}\rho_{N_{0}}^{(n,m)}+P_{mix}\rho_{mix}+P_{n}\rho_{n}\label{eq:noonmix3-1-1-2}
\end{equation}
where $\rho_{N_{0}}=|\psi_{N_{0}}\rangle\langle\psi_{N_{0}}|$, $\rho_{mix}$
is a mixture of states $|n\rangle|m+N_{0}\rangle$ and $|n+N_{0}\rangle|m\rangle$,
and $\rho_{n}$ is a state that gives predictions different to $j_{z}=\pm N_{0}/2$.
Only the first term will contribute a nonzero value of $\langle a^{\dagger N_{0}}b^{N_{0}}\rangle$.
The first term can also include superpositions of the different $|\psi_{N_{0}}\rangle$
with different $n,m$ but evaluation of the moment $\langle a^{\dagger N_{0}}b^{N_{0}}\rangle$
will be the same as if the system were in a mixture of those states
(due to the orthogonality). The \emph{relevant} ($Re$) values of
$n$, $m$ such that probabilities are nonzero can be determined from
the measurements of mode number and we assume the sums only includes
those nonzero contributions. We note that the first term is written
as a mixture of the NOON-type states. In some cases, such a mixture
can be equivalent to (and therefore rewritten as) a classical mixture
$\rho_{mix}$, but the nonzero moment $\langle\hat{a}^{\dagger N}\hat{b}^{N}\rangle$
cannot arise in this case. The value of $\langle\hat{a}^{\dagger N}\hat{b}^{N}\rangle$
is zero for any $\rho_{mix}$, and the prediction for $\langle\hat{a}^{\dagger N}\hat{b}^{N}\rangle$
given by $\rho$ is \textcolor{green}{} 
\begin{eqnarray}
|\langle\hat{a}^{\dagger N_{0}}\hat{b}^{N_{0}}\rangle| & = & |\sum_{n,m}a_{N_{0}}^{(n.m)}b_{N_{0}}^{(n,m)*}P_{N_{0}}^{(n,m)}\nonumber \\
 &  & \times\sqrt{\frac{(m+N_{0})!}{m!}}\sqrt{\frac{(n+N_{0})!}{n!}}|\nonumber \\
 & \leq & S\sum_{n,m}|a_{N_{0}}^{(n,m)}b_{N_{0}}^{(n,m)*}P_{N_{0}}^{(n,m)}|\label{eq:ineqcatfid-2-1-1}
\end{eqnarray}
We have used the prediction for $\langle a^{\dagger N_{0}}b^{N_{0}}\rangle$
for the state $|\psi_{N_{0}}\rangle$ and the definitions of $S$
as in the main text. \textcolor{black}{The measurement of the moment
$\langle\hat{a}^{\dagger N_{0}}\hat{b}^{N_{0}}\rangle$ thus allows
the determination of a lower bound on an effective fidelity for the
Schrodinger cat NOON state.} 

\emph{Correction term:} Now we consider that the experimentalist can
only confirm that the total probability of the ``nonrelevant'' ($NRe$)
outcomes is less than or equal to $\epsilon$. The contribution of
the ``nonrelevant'' terms to the $C_{N_{0}}$ (the sum of the $N_{0}$-th
order coherences) is bounded by the probabilities. For any density
matrix, the off-diagonal elements are bounded by the diagonal elements
that give the probabilities: Always $a_{N_{0}}^{(n.m)}b_{N_{0}}^{(n,m)*}\leq\frac{1}{2}$
and assuming $\sum_{NRe}P_{N_{0}}^{(n,m)}\leq\epsilon$, we find
\[
\sum_{NRe}a_{N_{0}}^{(n.m)}b_{N_{0}}^{(n,m)*}P_{N_{0}}^{(n,m)}\leq\epsilon/2
\]
Using that $\frac{(n+N_{0})!}{n!}\leq(n+N_{0})^{N_{0}}$, this implies
\begin{eqnarray}
\langle\hat{a}^{\dagger N_{0}}\hat{b}^{N_{0}}\rangle & \leq & \sum_{Re}a_{N_{0}}^{(n.m)}b_{N_{0}}^{(n,m)*}P_{N_{0}}^{(n,m)}\nonumber \\
 &  & \times\sqrt{\frac{(m+N_{0})!}{m!}}\sqrt{\frac{(n+N_{0})!}{n!}}\nonumber \\
 &  & +\sum_{NRe}a_{N_{0}}^{(n.m)}b_{N_{0}}^{(n,m)*}P_{N_{0}}^{(n,m)}(N_{up}+N_{0})^{N_{0}}\nonumber \\
 & \leq & S\sum_{n,m}|a_{N_{0}}^{(n,m)}b_{N_{0}}^{(n,m)*}P_{N_{0}}^{(n,m)}|+\frac{\epsilon}{2}(N_{up}+N_{0})^{N_{0}}\nonumber \\
\end{eqnarray}
Thus we know that 
\begin{eqnarray}
C_{N_{0}} & \geq & \sum_{Re}a_{N_{0}}^{(n.m)}b_{N_{0}}^{(n,m)*}P_{N_{0}}^{(n,m)}\nonumber \\
 & \geq & [\langle\hat{a}^{\dagger N_{0}}\hat{b}^{N_{0}}\rangle-\frac{\epsilon}{2}(N_{up}+N_{0})^{N_{0}}]/S\label{eq:cor}
\end{eqnarray}
where $N_{up}$ is the upper bound for the mode numbers, given that
the system cannot have infinite mode or particle (atom) number. For
the cases of interest to us on this paper, the total mode number is
the atom number $N$, which is fixed.

\section{Evaluation of Normalisation \label{sec:Appendix_Calculationnorm}}

We consider the state 
\begin{eqnarray}
|out\rangle & = & \sum_{m=0}^{N}d_{m}|m\rangle_{a}|N-m\rangle_{b}\,,\label{eq:bsoutputstate-1-1-2-1-1-1}
\end{eqnarray}
We quantify the $n$-th order quantum coherence by the parameter $C_{n}$
(that we have also called the catness-fidelity)

\textcolor{black}{
\begin{equation}
C_{n}=\mathcal{N}_{n,N}\sum_{m=0}^{N-n}|d_{m}d_{m+n}^{*}|\label{eq:catfidn-1}
\end{equation}
where $\mathcal{N}_{n,N}$ is a normalisation constant to ensure the
maximum value of $C_{n}$ is $1$. The normalisation $\mathcal{N}_{n,N}$
is determined by the bounds on the coherences of the density matrix
for a pure state. For example, where $n=N$, the maximum $|d_{0}d_{N}^{*}|$
is obtained for $d_{0}=d_{N}=\frac{1}{\sqrt{2}}$ with all other amplitudes
zero. Hence $|d_{0}d_{N}^{*}|\leq1/2$ and $\mathcal{N}_{N,N}=2$.}\textcolor{red}{{}
}\textcolor{blue}{}\textcolor{red}{{} }\textcolor{black}{Similarly,
for $n=N-1$ and $N\geq3$ (so that the $d$ terms in $d_{0}d_{N-1}^{*}+d_{1}d_{N}^{*}$
are all different), we find $d_{0}d_{N-1}^{*}+d_{1}d_{N}^{*}\leq1/2$
where in this case the maximum $\sum_{m=0}^{N-n}|d_{m}d_{m+n}^{*}|$
is found taking $d_{0}=d_{N-1}=d_{1}=d_{N}=\frac{1}{2}$. The maximum
value for more general $n$ and $N$ can be found numerically.}\textcolor{blue}{}

\textcolor{black}{(1) We start by analyzing $n=N$.} Then $C_{n}={\cal N}_{n,N}\vert d_{0}d_{N}^{*}\vert$.
There is only one term in the sum and therefore only two amplitudes
contributing to the sum. The number of terms is independent of $N$.
We can show that the maximum value of the sum of the coherences (namely
$\sum_{m=0}^{N-n}|d_{m}d_{m+n}^{*}|$) is given when $d_{0}=d_{N}=\frac{1}{\sqrt{2}}$,
and all other amplitudes zeros. Hence $C_{n}\leq{\cal N}_{n,N}\vert d_{0}d_{N}^{*}\vert={\cal N}_{n,N}\frac{1}{2}$
and the optimal normalisation is $ $${\cal N}_{n,N}=2$.\textcolor{blue}{{}
}\textcolor{red}{}\textcolor{blue}{}

(2) Next we consider $n>N/2$. Here $\sum_{m=0}^{N-n}|d_{m}d_{m+n}^{*}|=d_{0}d_{n}+d_{1}d_{n+1}+..d_{N-n}d_{N}$
and since $n>N-n$ the terms in the summation involve different $d_{i}$'s
which can be therefore be chosen independently apart from normalisation
requirements. Taking the $2(N-n+1)$ contributing amplitudes as equal,
and all other as zero, $\sum_{m=0}^{N-n}|d_{m}d_{m+n}^{*}|=\frac{(N-n+1)}{2(N-n+1)}=\frac{1}{2}$
which we verify is the maximum value. \textcolor{red}{}\textcolor{blue}{}

(3) \textcolor{red}{}For the remaining values, we determine the bounds
numerically. We analyse all these cases and fit an expression for
the maximum value of $\sum_{m=0}^{\frac{N}{2}}\vert d_{m}d_{m+\frac{N}{2}}^{*}\vert$.
On numerically analysing the cases $n<N/2$, we find that to a good
approximation $\sum_{m=0}^{N-n}\vert d_{m}d_{m+n}^{*}\vert\leq\cos\left(\frac{\pi}{[N/n]+2}\right)$,
where $[N/n]$ denotes the integer part of $N/n$ and hence ${\cal N}_{n,N}=1/\cos\left(\frac{\pi}{[N/n]+2}\right).$
We numerically verified this bound for all $N$ up to $500$. \textcolor{blue}{}\textcolor{black}{{}
}\textcolor{blue}{}

\section{Evaluation of $\langle\left(\hat{a}^{\dagger}\hat{b}\right)^{3}\rangle$
\label{sec:Appendix_CalculationMomentsN3}}

For $N=3$, we would like to measure the expectation value of the
following observable
\begin{eqnarray*}
\left(\hat{a}^{\dagger}\hat{b}\right)^{3} & = & \left(J_{x}+iJ_{y}\right)^{3}\\
 & = & J_{x}^{3}-iJ_{y}^{3}+i\left(J_{x}J_{y}J_{x}+J_{y}J_{x}^{2}+J_{x}^{2}J_{y}\right)\\
 &  & -\left(J_{y}^{2}J_{x}+J_{x}J_{y}^{2}+J_{y}J_{x}J_{y}\right).
\end{eqnarray*}
In the expansion, we have dropped the ``hats'' and used lower case
$x$ and $y$ in the subscripts of the $\hat{J}_{X}$ and $\hat{J}_{Y}$
defined in (\ref{eq:schspinab-1-1}) to simplify notation. The first
and second terms can be measured in experiments. However, we need
to express $J_{x}J_{y}J_{x}+J_{y}J_{x}^{2}+J_{x}^{2}J_{y}$ and $J_{y}^{2}J_{x}+J_{x}J_{y}^{2}+J_{y}J_{x}J_{y}$
in terms of some other measurements that can be carried out in experiments.
To this end, we define a rotated Schwinger operators as follows:
\begin{eqnarray*}
J_{\theta} & = & J_{x}\cos\theta+J_{y}\sin\theta\\
J_{\theta+\frac{\pi}{2}} & \equiv & G_{\theta}\\
 & = & J_{x}\cos\left(\theta+\frac{\pi}{2}\right)+J_{y}\sin\left(\theta+\frac{\pi}{2}\right)\\
 & = & -J_{x}\sin\theta+J_{y}\cos\theta.
\end{eqnarray*}
For $\theta=\frac{\pi}{4}$, these rotated operators correspond to:
\begin{eqnarray*}
J_{\frac{\pi}{4}}^{3} & = & \frac{1}{\sqrt{2^{3}}}\left[\left(J_{y}^{2}J_{x}+J_{x}J_{y}^{2}+J_{y}J_{x}J_{y}\right)\right.\\
 &  & \left.+\left(J_{x}J_{y}J_{x}+J_{y}J_{x}^{2}+J_{x}^{2}J_{y}\right)+J_{x}^{3}+J_{y}^{3}\right]\\
G_{\frac{\pi}{4}}^{3} & = & \frac{1}{\sqrt{2^{3}}}\left[-\left(J_{y}^{2}J_{x}+J_{x}J_{y}^{2}+J_{y}J_{x}J_{y}\right)\right.\\
 &  & \left.+\left(J_{x}J_{y}J_{x}+J_{y}J_{x}^{2}+J_{x}^{2}J_{y}\right)-J_{x}^{3}+J_{y}^{3}\right].
\end{eqnarray*}
Thus after manipulation  we obtain
\begin{eqnarray*}
J_{y}^{2}J_{x}+J_{x}J_{y}^{2}+J_{y}J_{x}J_{y} & = & \sqrt{2}\left(J_{\frac{\pi}{4}}^{3}-G_{\frac{\pi}{4}}^{3}\right)-J_{x}^{3}\\
\\
J_{x}J_{y}J_{x}+J_{y}J_{x}^{2}+J_{x}^{2}J_{y} & = & \sqrt{2}\left(J_{\frac{\pi}{4}}^{3}+G_{\frac{\pi}{4}}^{3}\right)-J_{y}^{3}
\end{eqnarray*}
Using the above expressions, we can then rewrite the moment $\left(a^{\dagger}b\right)^{3}$
in terms of the rotated Schwinger operators as:
\begin{eqnarray*}
\left(\hat{a}^{\dagger}\hat{b}\right)^{3} & = & J_{x}^{3}-iJ_{y}^{3}+i\left(J_{x}J_{y}J_{x}+J_{y}J_{x}^{2}+J_{x}^{2}J_{y}\right)\\
 &  & -\left(J_{y}^{2}J_{x}+J_{x}J_{y}^{2}+J_{y}J_{x}J_{y}\right)\\
 & = & J_{x}^{3}-iJ_{y}^{3}+i\left[\sqrt{2}\left(J_{\frac{\pi}{4}}^{3}+G_{\frac{\pi}{4}}^{3}\right)-J_{y}^{3}\right]\\
 &  & -\left[\sqrt{2}\left(J_{\frac{\pi}{4}}^{3}-G_{\frac{\pi}{4}}^{3}\right)-J_{x}^{3}\right]\\
 & = & 2J_{x}^{3}-2iJ_{y}^{3}+\sqrt{2}i\left(J_{\frac{\pi}{4}}^{3}+G_{\frac{\pi}{4}}^{3}\right)\\
 &  & -\sqrt{2}\left(J_{\frac{\pi}{4}}^{3}-G_{\frac{\pi}{4}}^{3}\right).
\end{eqnarray*}
which leads to the required result.

\end{document}